\documentclass[useAMS,usenatbib,referee]{mn2e}
\usepackage[dvips]{graphicx}

\title[The Progenitors of Core-Collapse Supernovae.]{The Progenitors of Core-Collapse Supernovae.}
\author[J. J. Eldridge and C. A. Tout]{J.J. Eldridge \thanks{E-mail: jje@ast.cam.ac.uk} and C.A. Tout \\ Institute of Astronomy, Madingly Road, Cambridge, CB3 0HA, England}

\pagerange{\pageref{firstpage}--\pageref{lastpage}} \pubyear{2004}
\begin{document}
\maketitle
\label{firstpage}

\begin{abstract}
We present maps of the nature of single star progenitors of supernovae and their remnants in mass and metallicity space. We find our results are similar to others but we have gone further in varying the amount of mixing and using various mass-loss schemes to see how the maps change. We find that extra-mixing, in the form of convective overshooting, moves boundaries such as the minimum mass for a supernova or WR star to lower masses. We also find that the pre-WR mass-loss determines the shape of our maps. We find that different mass-loss rates lead to quite different results. We find that the rise in luminosity at 2nd dredge-up places quite tight constraints on the masses of some progenitors and in particular the progenitor of supernova 2003gd.
\end{abstract}

\begin{keywords}
stars: evolution -- stars: supernovae progenitors -- stars: Wolf-Rayet
\end{keywords}

\section{Introduction.}
Supernovae (SNe) have most probably been observed since the dawn of astronomy. The earliest known surviving record of a supernova was made in the eleventh century by ancient Chinese astronomers (see \citet{SG02} for the details of these early observations), but the modern study of SNe began with Baade and Zwicky in the 1930's when they realised SNe are more luminous and rarer than the more common novae. The high luminosities and broad spectral lines led them to conclude that SNe are extreme explosions, the result of a dying star collapsing to a neutron star. This basic model still holds. Sixty-five~years of work has vastly increased our understanding but many questions about SNe remain unanswered.

A core-collapse SN occurs once a star has a core (usually of iron group elements) that cannot be supported by any further nuclear fusion reactions or electron degeneracy pressure. This leads to the core collapsing to a neutron star or black hole. In the process it releases a large flux of neutrinos that interact weakly with the surrounding stellar envelope depositing a large amount of energy ($10^{44} {\rm J}$) that can drive the star to explode. In massive cores supported by electron degeneracy pressure core-collapse can be initiated by electron capture onto nuclei.

Important questions are, `Which stars give rise to which SNe and leave which type of compact remnant?' The answers will provide a test of stellar evolution theory, give information for galactic chemical evolution and predict SNe birth rates and remnant populations. To solve these questions we must use our understanding of stellar evolution to model stars up to the point at which they become SNe. Three main factors affect the answer, the initial mass and metallicity of the star and mass-loss (or mass-gain in some binary systems) during its lifetime. \citet{H03}, using the models described in \citet{WHW02}, have considered these factors and answered some of our questions. While their study is informative, it lacks an accurate metallicity scale with only three metallicities studied $Z/Z_{\odot}=0$, $10^{-4}$ and $1$. They only investigate a single mass-loss scheme, do not investigate the lowest mass for a star to become a SN nor describe the SN progenitors. These factors limit comparison with observations. In this paper we expand upon their work and add details on the progenitors of SNe. These details are important for surveys of progenitors now underway by various groups, \citet{S03a} and \citet{VD03}.

From the literature we see that there are a variety of mass-loss schemes in use. We have compared a number of these and present a useful parameterisation. Within our study we have found many regions of interest, Wolf-Rayet (WR) stars, red giants (RG) and asymptotic giant branch (AGB) stars. Their relative abundance, dependent on the initial mass function, and luminosity are of key interest as well as whether they do achieve the conditions to undergo core collapse. We have compared our results with observations of the known progenitors to date. In a few years there will hopefully be enough known progenitors to be able to fully test our predictions.

\section{Construction of Models.}

\subsection{Stellar Evolution Code Details.}
We have used the Cambridge STARS stellar evolution code originally developed by \citet{E71} and updated most recently by \citet{P95} and \citet{E03}. Further details can be found at the code's home page (http://www.ast.cam.ac.uk/$\sim$stars). In this work we use 21 zero-age main-sequence models that have masses from $5$ to $200M_{\odot}$ with a uniform composition determined by $X=0.75-2.5Z$ and $Y=0.25+1.5Z$, where $X$ is the mass fraction of hydrogen, $Y$ that of helium and $Z$ is the initial metallicity. It takes values from $10^{-5}$ to $0.05$, equivalent to $\frac{1}{2000}Z_{\odot}$ to $2\frac{1}{2}Z_{\odot}$. The masses and metallicities used are shown on the axes of Figure 1. The models are evolved with a mixing length parameter $\alpha=2.0$ in the formalism of \citet{BV58}. This value gives a good fit to a realistic solar model \citep{P98}.

We have also created maps with alternate compositions: $X=0.7$, $Y=0.3-Z$ and $X=0.76-3Z$, $Y=0.24+2Z$. With constant hydrogen abundance the variation with metallicity is accentuated. For example 2nd dredge-up occurs at higher masses with increasing metallicity than we find below. Changing the primordial hydrogen abundance from 0.75 to 0.76 as in the second case only produces small changes to our results.

\subsubsection{Nuclear Reactions and Opacities.}
At the time of \citet{P95} only six composition variables were modelled. We have extended this to eight: H, He, C, N, O, Ne, Mg and Si. We have also updated the reaction rates according to the NACRE collaboration \citep{NACRE}, replacing a number of those originally taken from \citet{CF1988}. We have further updated the reaction network to follow the later stages of burning although we encounter numerical difficulties at the onset of neon and oxygen burning because these reactions occur at a much more rapid pace than the previous evolution with timescales of a year and below. During these burning stages the envelope is only affected to a small degree since $t_{\rm nuclear} << t_{\rm thermal}$. Therefore we use models after core carbon burning to estimate the details of progenitors. We have confirmed from models that do experience a neon burning phase that the late stages of evolution are rapid and have a negligible effect on the final observable state of the progenitor.

We use the updated tables of \citet{E03} that accurately follow the opacity variation owing to the He/C/O ratio changes during evolution by including all the tables available from the OPAL project \citep{IR96}. We found important differences in the structure of Wolf-Rayet (WR) stars and so we are able to model these stars with increased accuracy.

\subsubsection{Convective Overshooting.}
Convective mixing is modelled simultaneously with the structure with a diffusion equation that allows us to correctly follow stable semi-convection. Our convective overshooting is not parameterised by a fixed fraction of the pressure scale height, $H_{\rm P}$. Instead we modify the classical Schwarzschild criterion for instability, normally expressed as
\begin{equation}
\nabla_{{\rm r}} > \nabla_{{\rm a}},
\end{equation}
for convective instability to occur, where $\nabla_{{\rm r}}$ and $\nabla_{{\rm a}}$ are the radiative and adiabatic gradients respectively. To include overshooting we alter this condition to
\begin{equation}
\nabla_{r} > \nabla_{a} - \delta,
\end{equation}
where
\begin{equation}
\delta = \frac{\delta_{{\rm ov}}}{2.5+20\zeta+16\zeta^{2}},
\end{equation}
$\zeta$ is the ratio of radiation to gas pressure and $\delta_{{\rm ov}}$ is the overshooting parameter that we vary to include convective overshooting or to remove it. Further details of the inclusion of this scheme in the Eggleton code can be found in \citet{SPE97}. They found from fitting models to eclipsing binaries $\delta_{{\rm ov}}$ to be  $0.12$. A similar trend was confirmed with a large pool of binaries in \citet{P97}. This leads to overshooting lengths $l_{{\rm ov}}$ between $0.25$ and $0.32 H_{\rm P}$ in stars of mass range $2.5$ to $6.5 M_{\sun}$ for more massive stars the overshooting length remains around $0.3 H_{\rm P}$. Another way of quantifying the effect of overshooting is to measure the mass of the convective core of a ZAMS model. For the $6.5 M_{\odot}$ star the mass over which mixing occurs grows from $1.6$ to $2.1 M_{\odot}$, for a $10M_{\odot}$ star the growth is from $3.2$ to $4.1 M_{\odot}$ and for a $100M_{\odot}$ star from $83M_{\odot}$ to $87M_{\odot}$.

In our calculations we use $\delta_{{\rm ov}}=0.12$ to include convective overshooting and set it to zero to remove overshooting. The amount of overshooting affects the core mass during evolution. This affects the nature of the remnants and also the mass of stars that undergo second-dredge up. The extra mixing owing to convective overshooting can be considered to mimic any mixing process such as rotation or gravity wave mixing, not just convective overshooting. For a review of the various extra mixing processes and the possible anomalies of stellar evolution they could resolve see \citet{b7b}.

\subsection{Mass-Loss Prescriptions.}

Mass-loss modifies the evolution of a star by affecting the surface conditions and the total mass of the star. While theoretical mass-loss rates do exist (Vink, de Koter \& Lamers 2000, 2001) it is sensible to use empirical mass-loss rates if we aim to model what is observed. We have tested the theoretical rates of Vink et al. and found the results agree well with those with empirical mass-loss rates. We divide our mass-loss rates into two main categories those for Wolf-Rayet (WR) stars, that have lost their hydrogen envelope, and pre-WR or hydrogen-rich evolution. The nature of mass loss changes once the hydrogen envelope has been removed and observations show quite different mass-loss rates. WR stars have greater mass loss than an OB star of the same luminosity. The mechanism for mass loss in OB stars is radiatively driven winds. The WR star mechanism is not known for certain. It could be driven by radiation, pulsations or opacity \citep{HL96}.

Our resultant maps are strongly dependent on the mass-loss prescription. We have made a number of grids using prescriptions similar to \citet{MM}, \citet{H03} and \citet{DT03}. For our main maps we have adopted the prescription below, they are essentially the NL rates from \citet{DT03} but with a modification at low metallicity ($\le 10^{-4}$) as detailed below. In section 5.1 we present detailed comparison of the different mass-loss prescriptions we have tested.

\subsubsection{Pre-WR Evolution.}
We use the mass-loss rates of \citet{dJ}, here after JNH, except for a small region at low metallicity and high stellar masses detailed below. The JNH rates are old and complex but considered to be the most accurate \citep{Crow2001}. The rates are dependent on surface luminosity and temperature and are derived from a large pool of observations. We apply the commonly adopted scaling with metallicity of $\dot{M}(Z) = \dot{M}(Z_{\odot}) \times (Z/Z_{\odot})^{0.5}$ as in \citet{H03} and \citet{DrayThesis}. This scaling arises from the assumption that stellar winds are line driven and with lower surface opacity at lower metallicity there are weaker winds. However while there is agreement that mass-loss scales in this form there is a range of suggested values for the exponent. We have tested the sensitivity to changes in this scaling by producing a grid with $(Z/Z_{\odot})^{0.6}$ and higher values for the exponent (up to 0.69 suggested by Vink et al. As expected a larger exponent increases variation with metallicity: the type II/Ibc SN boundary varies with metallicity to a greater degree. Previous authors have commonly assumed that this scaling applies at all stages of evolution even though there is some evidence that it varies with spectral type. This seems likely because the exponent assumes the winds are line driven while mass loss from giants is driven by some quite different unestablished mechanism. If we do not scale the mass loss for giant stars ($T_{\rm eff} <3.7$) the result is that the minimum initial mass for WR star formation decreases with decreasing metallicity. This is contrary to other results. In another test we used an exponent of 0.5 for all stars except hot OB stars ($T > 10^{4}$) where we used an exponent of 0.7. This produces results very similar to those of our preferred results using 0.5 for all stars.

\begin{figure}
\includegraphics[height=80mm]{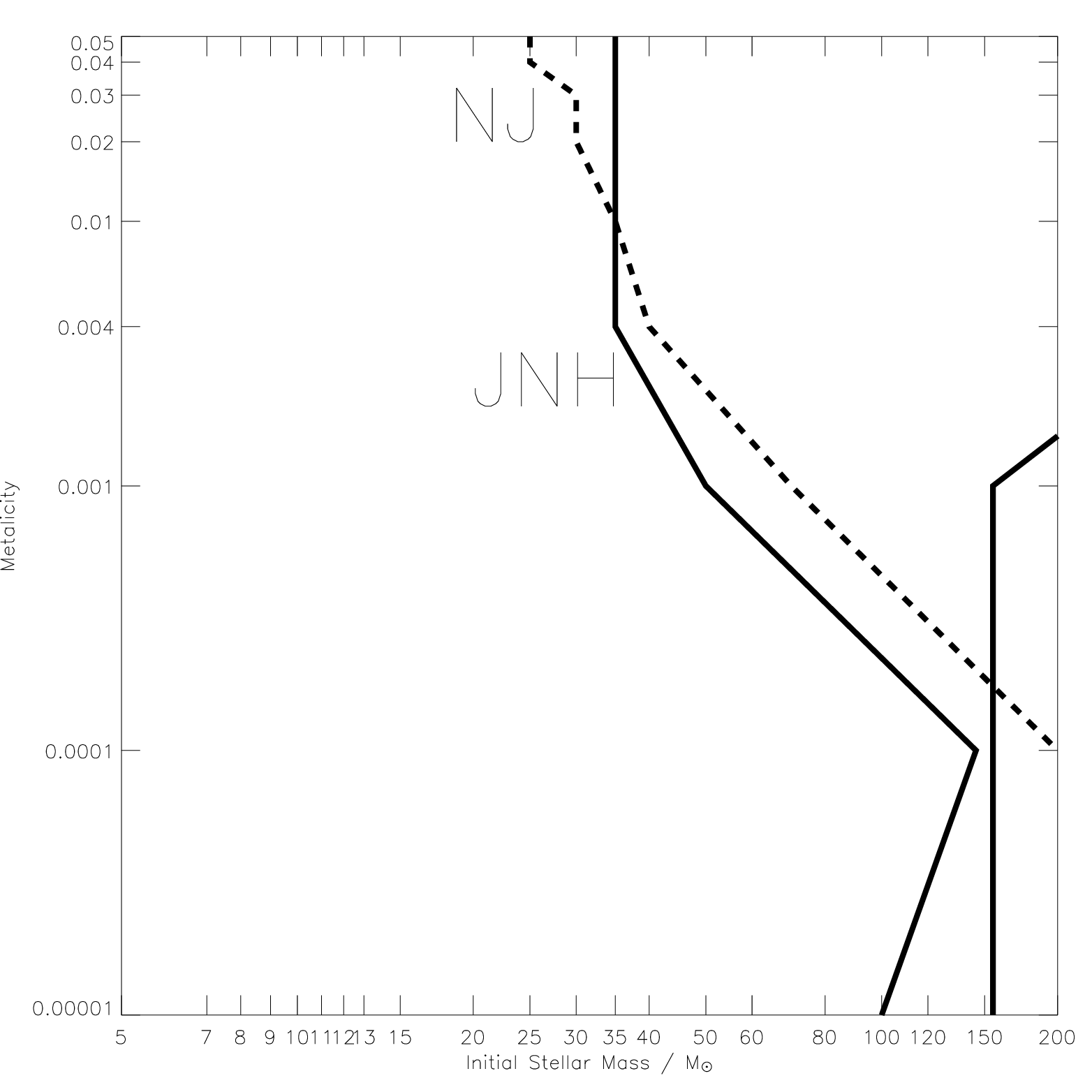}
\caption{Borders between SNe Type II/Ibc. Dashed line for NJ rates without convective overshooting, solid line for JNH rates without convective overshooting. See Table \ref{tml} for description of rates. Generally the stars to the right of the lines are type Ibc SNe, those to the left are type II SNe. Although at $Z \le 0.001$ for JNH rates stars to the right of the second line are type II SNe. The metallicities and masses or our models are used as labels on the axes.}
\label{sneborders}
\end{figure}

We encounter problems scaling the JNH rates when $Z \le 10^{-4}$ for the highest mass stars ($M \ge 150M_{\odot}$ when $Z=10^{-4}$ and $M \ge 80M_{\odot}$ when $Z=10^{-5}$). The stellar parameters are outside of the region covered by JNH. Using simple extrapolation the mass-loss rate drops to very small values and leads to strange structure over the map. There is still a minimum mass for Ibc stars but some of the stars more massive than this minimum do not lose all their hydrogen before SNe. The problem can be seen in Figure \ref{sneborders} where with the JNH rates at low metallicity the minimum mass for Ibc SNe decreases with decreasing metallicity. This is because the nature of first dredge-up differs in these stars and complicates the mass-loss history. However, if we replace the JNH rates with those of \citet{NJ90} (NJ) which are based on the same set of observational data but are expressed more simply a smoother structure is obtain. All the stars above the minimum Ibc SN mass do undergo type Ibc SNe. We also find that none of our models at $Z=10^{-5}$ undergo type Ibc SNe. We do not use the theoretical rates of Vink et al. because they do not extend down to these low metallicities. We consider the rates of NJ rather than JNH to be the best extrapolation method at the current time at low metallicity and high initial mass. We have also experimented with the low metallicity theoretical mass-loss rates of \citet{KD2002} and find similar results.

\subsubsection{WR Evolution.}
WR evolution begins just before the hydrogen envelope is completely removed from the star. The stages of evolution run in the sequence of WN, WC and WO. In each case a particular element dominates in the emission spectrum, nitrogen, carbon or oxygen respectively. However the definition of when a model enters each stage is arbitrary. We use those defined in \citet{DrayThesis}.

\begin{itemize}
\item WN: When $X_{\rm Surface} < 0.4$ and $T_{\rm eff} > 10^{4}$.
\item WC: When $X<0.001$ and $0.03 < (X_{\rm C}+X_{\rm O})/Y < 1.0$.
\item WO: When $X<0.001$ and $(X_{\rm C}+X_{\rm O})/Y > 1.0$.
\end{itemize}

During WN and WC evolution we use the rates of \citet{NL00} that depend on luminosity and surface abundance They find that the mass-loss rate depends strongly on composition as well as luminosity. This dependence gives lower mass-loss rates than the mass-dependent rates of \citet{Langer} and the mass-dependant rates that \citet{NL00} also derive. We have tested both sets of mass-dependant rates and find that they lead to extreme mass loss from WR stars at and above solar metallicity and lead to the remnant becoming a white dwarf. Our preferred rates are therefore the main \citet{NL00} rates.

One straightforward test of this mass-loss prescription is to compare the number ratios of WR stars to O stars, and the WR subtype ratios. \citet{DT03} performed this test in detail and found good agreement to observed ratios. The one exception was the WN/WR ratio with the theoretical value being too low. There are a number of plausible reasons for this such as binaries affecting mass-loss and misclassification of some WN stars due to changing spectral features as metallicity changes.

\section{Supernovae Type Determination.}

There are two main types of SNe those without hydrogen in their spectra, type I, and those with hydrogen in their spectra, type II. Type I progenitors have lost their hydrogen envelopes and are white dwarfs or WR stars, while type II progenitors have retained their hydrogen envelopes. This gives us a basic method to discriminate between them.

\subsection{Type II}
For a star with a model at the end of core carbon burning we say it will become a type II SNe if there is any hydrogen in its envelope. Further more we know observationally that type IIP SNe have retained a good proportion of their original hydrogen. This hydrogen leads to the light curve following a plateau phase powered by a moving hydrogen ionization front. Type IIL SNe only retain a small fraction of their hydrogen as the light curve decays linearly. We adopt the value of \citet{H03} with type IIP SNe occurring until the mass of hydrogen in the star drops below $2.0 M_{\odot}$ when type IIL SNe occur.

\subsection{Type I}
Type I SNe are divided into types Ia, Ib and Ic. SNe Ia are extremely bright events thought to be the explosive carbon burning of a degenerate white dwarf that has reached the Chandrasekhar mass ($M_{\rm Ch}$) by accretion; this is not a core collapse event. However types Ib and Ic are core-collapse SNe in WR stars that have lost their hydrogen envelopes in a wind or by binary interaction. They are discriminated by the presence or absence of helium lines in the spectrum. However the differentiation between models that are type Ib or Ic is not obvious so we will label them as Ibc. All of our progenitors still retain some helium and the loss of this final layer is likely to depend on the very end phase of evolution and the explosion itself.

We also adopt from \citet{H03} a method of determining the strength of a type Ibc SN. This will allow us to quantify the number of SNe that have no observable display and are thus unseen. This allows us to estimate ratios of type II to type Ibc SNe. The largest progenitors are thought to give no display because the core is so massive that even with a large explosion energy nothing escapes the forming black hole. The exception would be if a jet driven SNe occured that makes a black hole and produces an observable display \citep{MWH01}. These ranges are taken from \citet{EW88} and are similar to those used for determining the supernovae remnant (see next section).
\begin{itemize}
\item $M_{\rm Core}({\rm He}) > 15M_{\odot}$, no display.
\item $15M_{\odot} > M_{\rm Core}({\rm He}) > 8M_{\odot}$, faint SN.
\item $8M_{\odot} > M_{\rm Core}({\rm He}) > 5M_{\odot}$, possibly faint SN.
\item $M_{\rm Core}({\rm He}) < 5M_{\odot}$, bright SN.
\end{itemize}

\section{Supernovae Remnant Determination.}
Determination of the remnant formed at the heart of a SN is a black art. The physics is extreme and poorly understood. However there are many prescriptions which rely on the conjecture that a more massive core leads to a more massive remnant. We consider here three types of stellar remnant: white dwarfs, neutron stars and black holes.

White dwarfs come in three flavours. They can be made of helium (a HeWD), a mixture of carbon and oxygen (a COWD) or a mixture of oxygen and neon with some sodium and magnesium (an ONeWD). HeWDs are formed by very low mass stars or more probably in binary stars and are not of interest to us here. COWDs are also not formed in SNe but are the remnant cores of stars that lose their envelopes on the asymptotic giant branch (AGB). At solar metallicity stars in the mass range $0.8$ to $7 M_{\odot}$ (no convective overshooting) produce a COWD while larger stars, up to $9M_{\odot}$ develop an ONe core before the AGB. Some of these lose their envelope and form an ONeWD. The most massive undergo a SN driven by electron capture on to magnesium. The dividing line between these two is of great interest but is somewhat uncertain and depends on the mass-loss rate and the details of the thermal pulses in AGB stars.

From our models we decide that stars with cores that grow by thermal pulses after 2nd dredge-up form COWDs or ONeWDs. This corresponds to a CO core mass $<1.38$ after 2nd dredge-up. All other stars undergo a SN and form a neutron star or black hole. To distinguish between neutron stars and black holes there are a number of different prescriptions. We compare that of \citet{H03} with ours.

\subsection{He Core Mass.}
\citet{H03} use a simple system based on the helium core mass at the end stage of evolution. This is useful because the helium burning life-time is slower than for the later stages of burning so the core mass increases by only a small amount in the late burning stages. However the method relies on setting the values for different remnants by using simulations of SNe. The values used are:
\begin{itemize}
\item Neutron Star: $M_{\rm Core}({\rm He}) < 8M_{\odot}$.
\item Black Hole (by fall back): $8M_{\odot} < M_{\rm Core}({\rm He}) < 15M_{\odot}$.
\item Black Hole (directly): $M_{\rm Core}({\rm He}) > 15M_{\odot}$.
\end{itemize}
These values are based on the SN models of \citet{Fry99}. By using this scheme we can compare directly with \citet{H03}. We should also note that SN that form a black hole directly also have no display. While those SN forming black holes by fall back will be faint. Unless a jet driven SNe occurs producing a black hole and an observable display.

\subsection{Core Mass by Energy.}
We use a core mass defined by the binding energy of the star. We assume a neutron star is formed at the centre of the star after core collapse of mass $M_{\rm Ch}=1.44M_{\odot}$. This produces about $10^{46} {\rm J}$ of energy from the release of gravitational binding energy in neutron star formation. We then assume a hundredth of this energy is transferred into the envelope by some unknown mechanism. The current suggestion is the transfer occurs via neutrinos released from forming the proto-neutron star that are thermalised within the envelope or dense outer parts of the core.

We integrate the binding energy of the star from the surface towards the centre until we reach $10^{44} J$. The envelope outside this region is ejected with the remaining amount forming the remnant.
\begin{equation}
\int^{M_{*}}_{M_{\rm rem}} \Big( U-\frac{GM}{r(M)} \Big) dM \, = \, 10^{44} {\rm J} 
\end{equation}

 If we have a $M_{\rm rem} > 2M_{\odot}$ it is a black hole. The advantage of this method is that we are able to estimate the remnant masses and ejected masses quickly for large numbers of models. The resulting minimum mass for a black hole is similar to that from the He core mass method. However these remnant masses are at best a lower bound as we use a very simple model for a complex event. 

\section{Results.}

\subsection{Comparing mass-loss schemes}

To demonstrate the effects of the main mass-loss prescriptions in use we evolve three stars at solar metallicity with six different mass-loss schemes. The prescriptions are detailed in Table \ref{tml}.

\begin{table}
\caption{The mass-loss prescriptions identifiers. JNH and NJ are as described in the main text, NL is from \citet{NL00} and L from \citet{Langer}.}
\label{tml}
\begin{tabular}{lcc}
\hline
Identifier & Pre-WR & WR \\
\hline
NL & JNH & NL\\
NN & NJ  & NL\\
$\frac{1}{2}$MM & JNH & L \\
MM & JNH$\times 2$ & L\\
HG & NJ  & L$\times \frac{1}{3}$\\
3HG & NJ  & L\\
\hline
\end{tabular}
\end{table} 

The MM scheme is as used by \citet{MM} and HG is as used by \citet{H03}. The numerical constants are extra factors which are used to enhance or suppress the mass-loss in certain stages of evolution. For example the rates of \citet{Langer} are reduced by $\frac{1}{3}$ due to clumping \citep{HKO98}. For the HG WR rates we apply the scaling of \citet{WL1999} with and without the reduction by one third. For pre-WR evolution JNH and NJ are based on the same set of observational data however the formulae given in the papers depend on 20 or 3 terms respectively. Thus JNH rates have much more structure over the HR diagram. For WR evolution NL rates are based on more recent data and depend on luminosity and surface composition while the L rates are older and depend only on mass of the star.

\begin{table}
\caption{Details of Models with different mass-loss prescriptions, $Z=0.02$. $M_{i}$ is the initial ZAMS mass, $L$ is the star's luminosity and $M_{\rm P}$ the mass of the progenitor before SN. }
\label{tres}
\begin{tabular}{lcccc}
\hline
                  & Mass-Loss  &                                  &         &        \\
$M_{i}/M_{\odot}$ & Identifier & $\log_{10}(\frac{L}{L_{\odot}})$ & SN Type & $M_{\rm P}/M_{\odot}$ \\
\hline
 25 & NL              & 5.53 & IIP & 16.0  \\
 25 & NN              & 5.40 & IIL & 9.73  \\
 25 & $\frac{1}{2}$MM & 5.53 & IIP & 16.0  \\
 25 & MM              & 5.50 & IIP & 11.0  \\
 25 & HG              & 5.70 & IIL & 9.85  \\
 25 & 3HG             & 5.70 & IIL & 9.85  \\
\hline
 40 & NL              & 5.36 & Ibc & 8.49  \\
 40 & NN              & 5.44 & Ibc & 9.43  \\
 40 & $\frac{1}{2}$MM & 5.17 & Ibc & 5.41  \\
 40 & MM              & 5.13 & Ibc & 4.90  \\  
 40 & HG              & 5.23 & Ibc & 7.02  \\
 40 & 3HG             & 4.89 & Ibc & 3.25  \\
\hline
120 & NL              & 5.31 & Ibc & 7.69  \\ 
120 & NN              & 5.20 & Ibc & 6.53  \\
120 & $\frac{1}{2}$MM & 5.15 & Ibc & 5.11  \\
120 & MM              & -   & none& $<1.4$  \\
120 & HG              & 5.13 & Ibc & 5.23  \\
120 & 3HG             & -   & none& $<1.4$  \\
\hline
\end{tabular}
\end{table} 

From Table \ref{tres} it is possible to see many trends. Doubling the pre-WR mass-loss rate generally leads to smaller remnants and less hydrogen in the envelope of the progenitor. Of special interest are the rates of \citet{H03} and \citet{MM}. In the sample above these schemes lead to smaller remnant masses and in the most extreme case a WD. For these models the evolution was halted once they reached the Chandrasakar mass. They occur because of high pre-WR mass-loss followed by mass dependent rates for WR evolution. At lower masses the agreement with other prescriptions is better, although the NN rates lead to a lower minimum initial mass to form a WR star.

The overall shapes of our maps are consistent between mass-loss prescriptions but specific values change between them. Of special interest is the comparison between NJ as in \citet{H03} and our preferred rates JNH. If we compare the two graphs in Figure \ref{sneborders} we see that the former lead to greater mass loss and therefore a lower mass for Ibc SNe at solar metallicity. Our minimum mass for type Ibc SNe with NJ rates is $4M_{\odot}$ lower at solar metallicity that that from \citet{H03} due to using the NL mass-loss rates for WR evolution. The NJ rates also produce greater variation with metallicity. This is because the NJ rates are extremely simple with only 3 terms in the formula for mass-loss while JNH employ 20 terms that can resolve greater structure over the HR diagram and produce more accurate models. The NJ scheme blurs out the details and makes the mass-loss too large at some points of evolution. Here we prefer to use the JNH mass-loss rates until at very low metallicities when the stellar parameters move outside the valid region of JNH where we use the NJ rates which can be extrapolated easily. 

From Table \ref{tres} it can be seen HG mass-loss rates lead to the production of lower mass progenitors. The result on our maps is to erode the structure in the upper right-hand corner. For example Figure \ref{mapD} with HG rates would have neutrons stars as the only possible remnants when $Z \ge 0.03$ and black holes only forming directly when $Z \le 0.004$.

\subsection{Progenitor, SN and SN Remnant Diagrams}

Our maps tend to broadly agree with those of \citet{H03}. Although, owing to our specific coverage in metallicity, we obtain more structure within the map. All of the structure is due to the mass-loss prescriptions we have used in producing these models. It is important to note however that while the broad nature of these maps are correct some of the fine structure would in practice be blurred out by different helium content, rotation, magnetic fields and other mixing processes. We assess our maps for each variable individually.

\begin{figure}
\includegraphics[height=80mm]{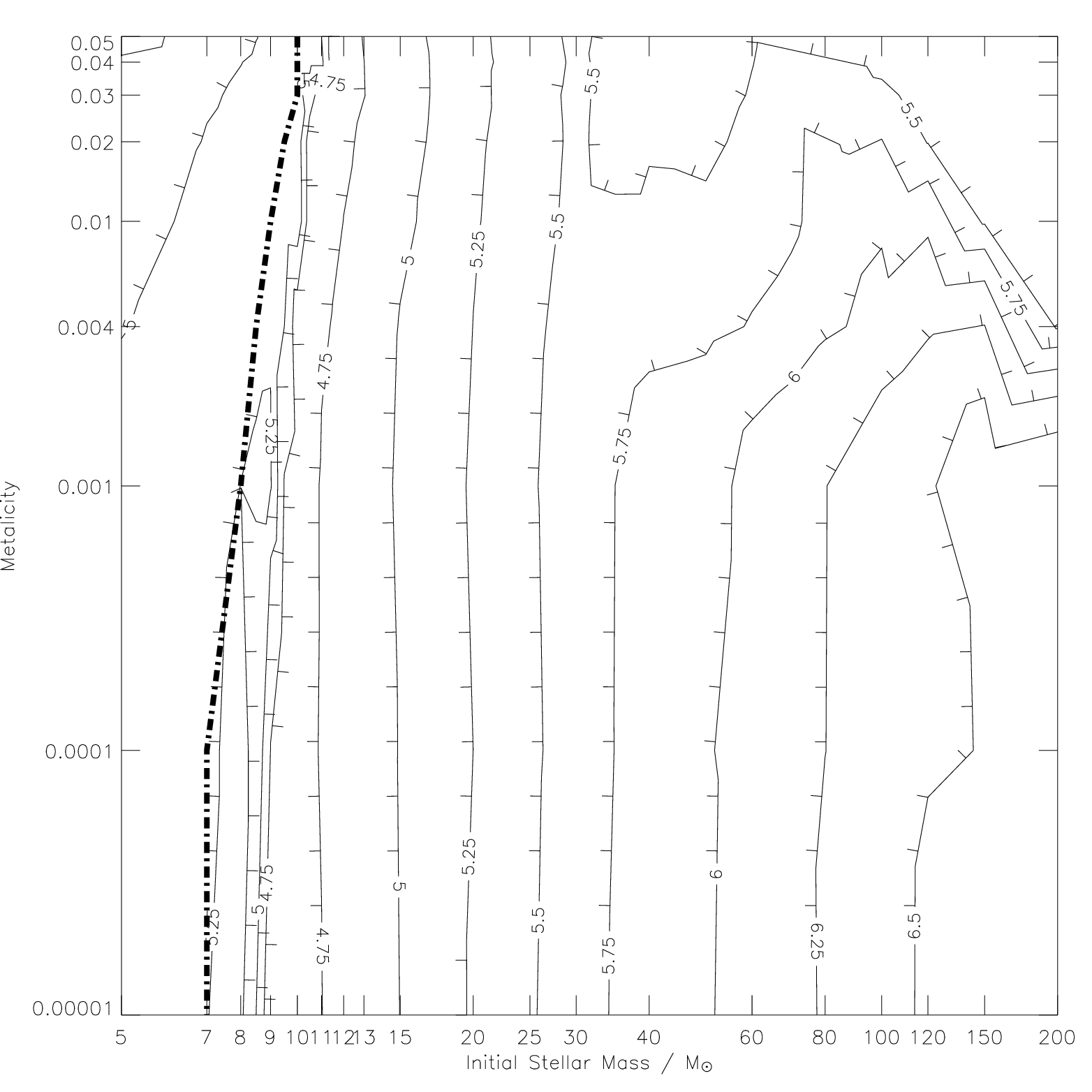}
\includegraphics[height=80mm]{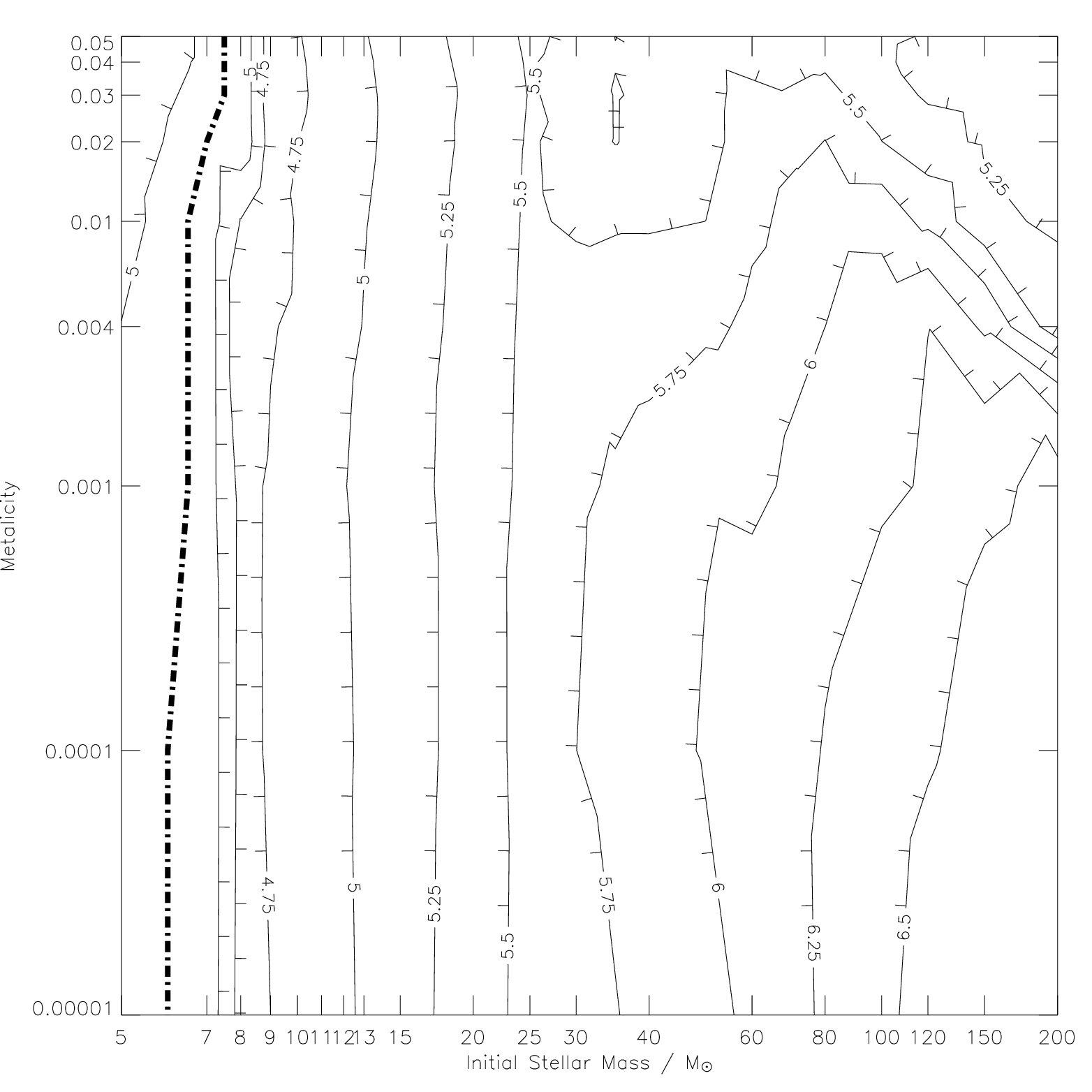}
\caption{Luminosity immediately before SN, left no-overshooting, right overshooting. Units on contours are $\log (L/L_{\odot})$. The dot-dashed line represents the minimum mass for a SNe.}
\label{mapA}
\end{figure}

\subsubsection{Luminosity, Figure \ref{mapA}}
Most important here is the 2nd dredge-up line on the left half of  the diagram its position changes with inclusion of overshooting. Without overshooting 2nd dredge-up occurs for zero-age masses between 8 and 11 $M_{\odot}$. With overshooting this range drops to between 7 and 9 $M_{\odot}$. Interestingly the most massive of these stars do form CO cores above $M_{\rm Ch}$ and go through carbon burning forming an ONe core. Then, before the surface convection zone penetrates the helium core, a convection zone forms between the hydrogen and helium burning shells that dredges carbon and oxygen into this intershell region thus decreasing the CO core mass. When the surface convection zone dredges up helium it also dredges up the carbon and oxygen. This behaviour therefore means that during 2nd dredge-up the surface carbon and oxygen abundances increase, rather than the normal decrease found in lower mass stars.

Another feature in the diagram is the structure in the top right hand corner. As mass increases we see the last few red giant progenitors are quite bright but are then replaced by WR stars and the pre-SN luminosity drops abruptly. The luminosity increases again as higher mass WR stars form. At the highest initial masses these WR stars once more become less massive and therefore less luminous. This structure becomes less pronounced at lower metallicities owing to the lower mass-loss rates.

\begin{figure}
\includegraphics[height=80mm]{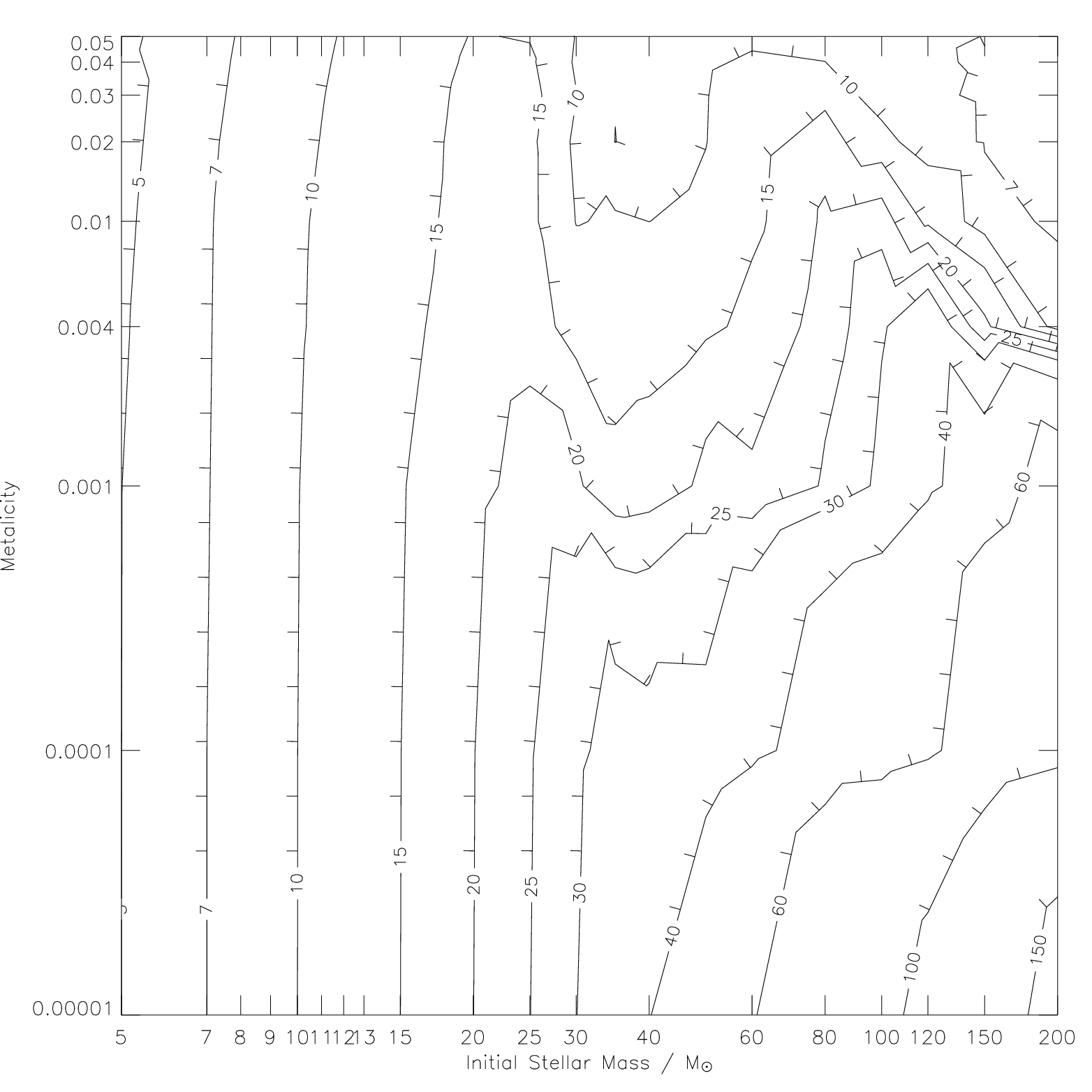}
\includegraphics[height=80mm]{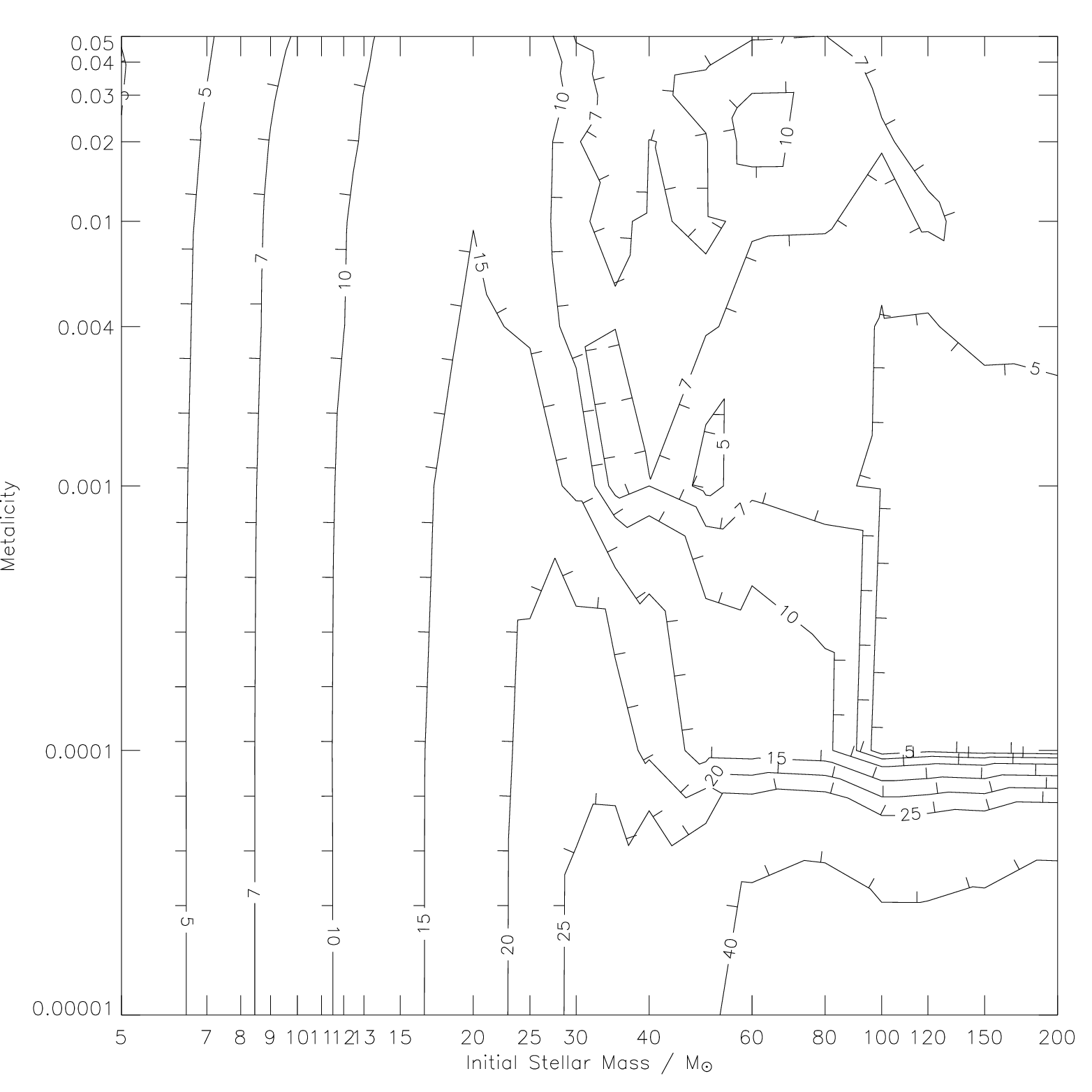}
\caption{Left final mass of progenitor before SN, right mass ejected in SN. Both are with overshooting, units on contours are mass in $M_{\odot}$.}
\label{mapB}
\end{figure}

\subsubsection{Final \& Ejected Mass, Figure \ref{mapB}}
The final mass diagram has very similar structure to the luminosity diagram and again reflects how mass loss is more severe in the high-metallicity region and drops off at low metallicity. The mass ejected by the SNe is also interesting because it demonstrates that a WR star ejects very little mass because it is more dense and so more tightly bound than a red giant progenitor.

\begin{figure}
\includegraphics[height=80mm]{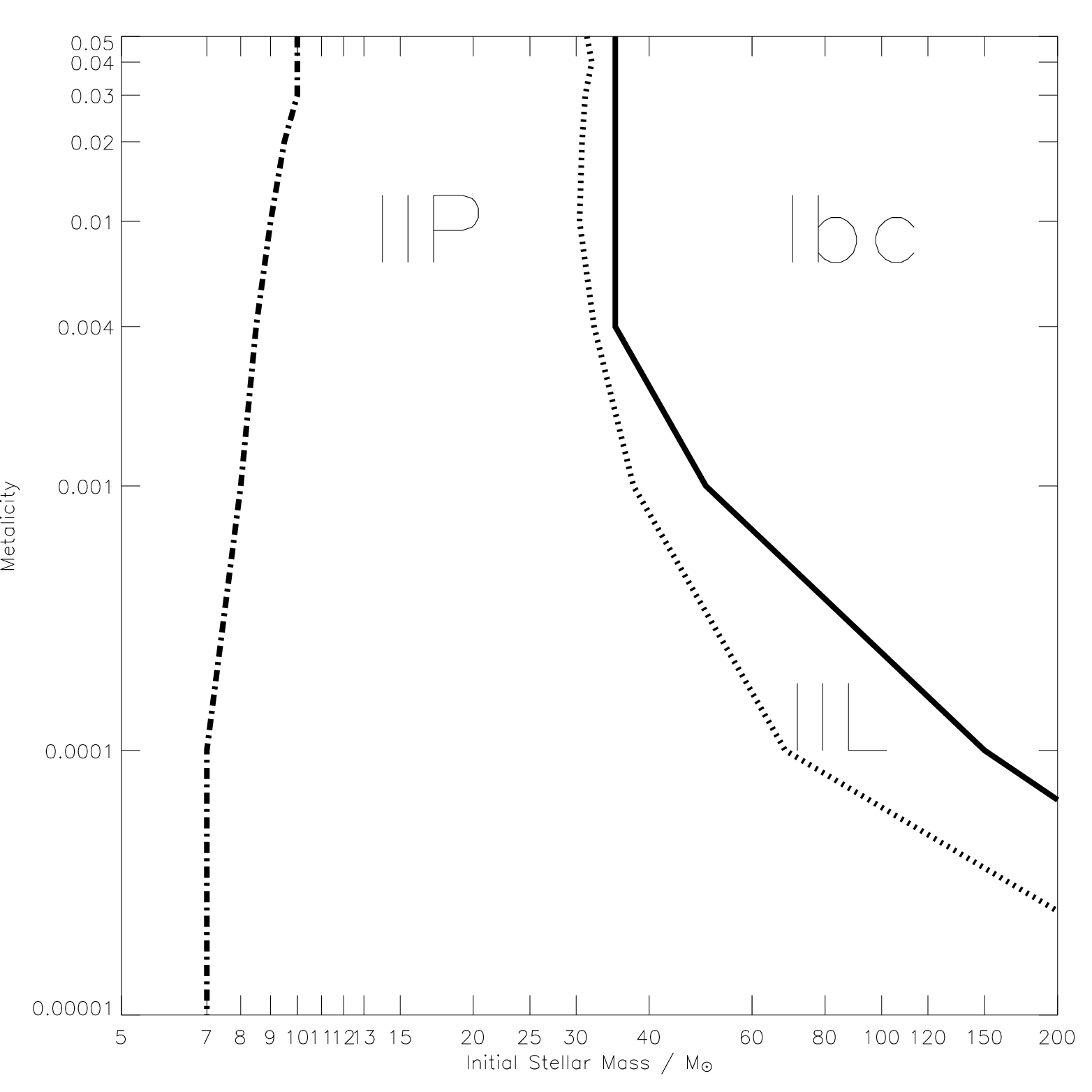}
\includegraphics[height=80mm]{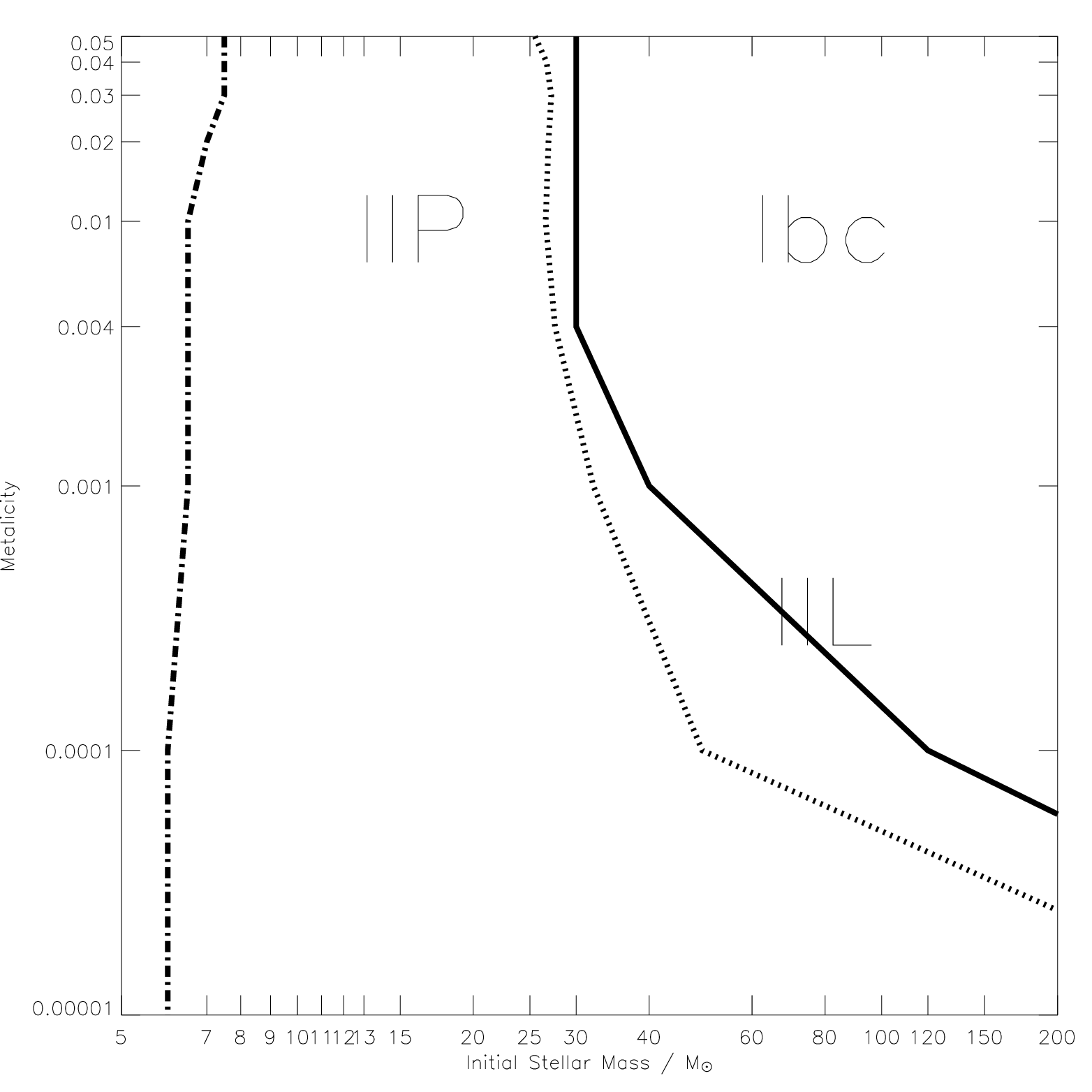}
\caption{SNe Type, left no-overshooting, right overshooting. Solid line - boundary between type II/I SNe, dotted line boundary between type IIP/IIL from total mass of hydrogen. The dashed-dotted line is the estimate of the location of the lowest progenitor mass for a SN.}
\label{mapC}
\end{figure}

\begin{figure}
\includegraphics[height=80mm]{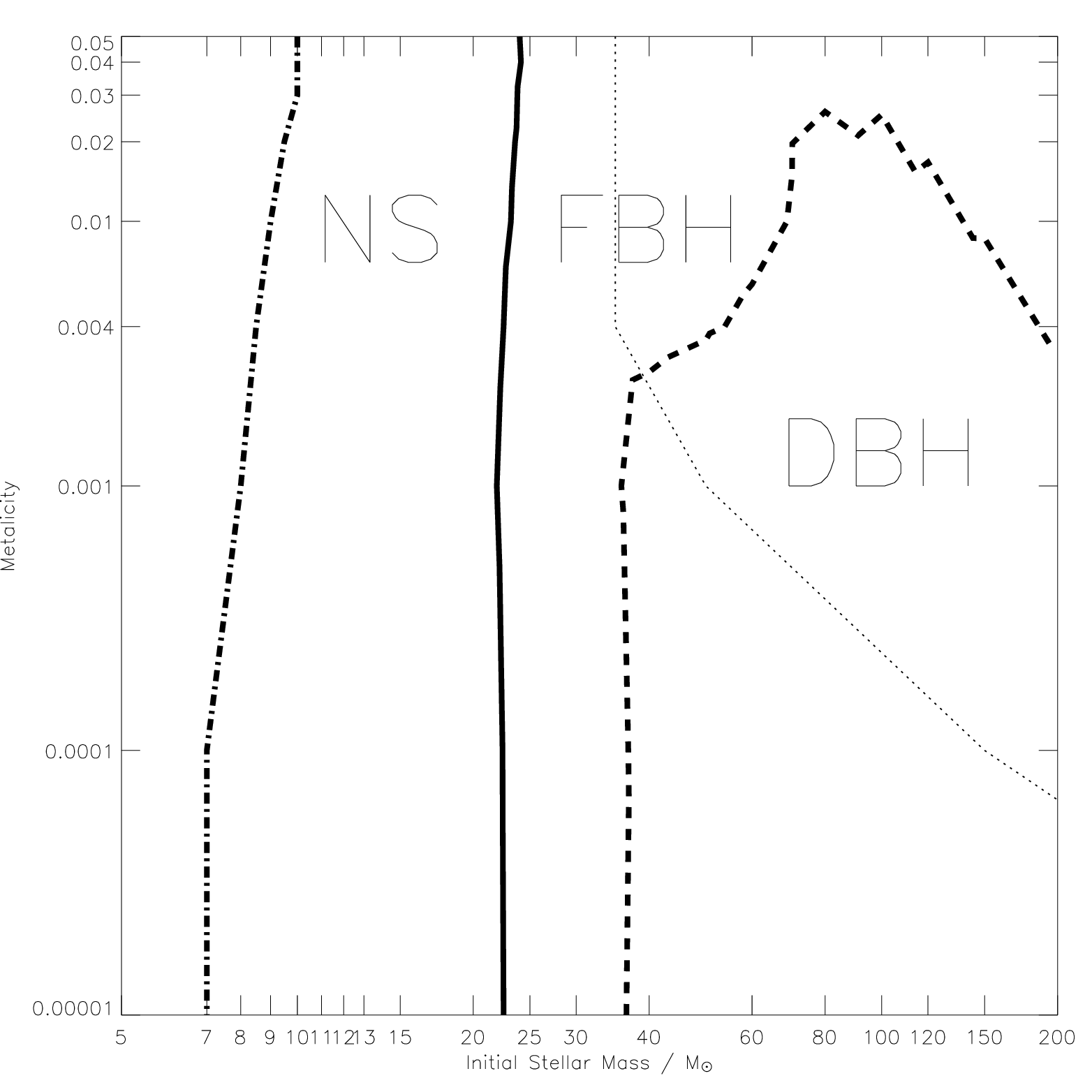}
\includegraphics[height=80mm]{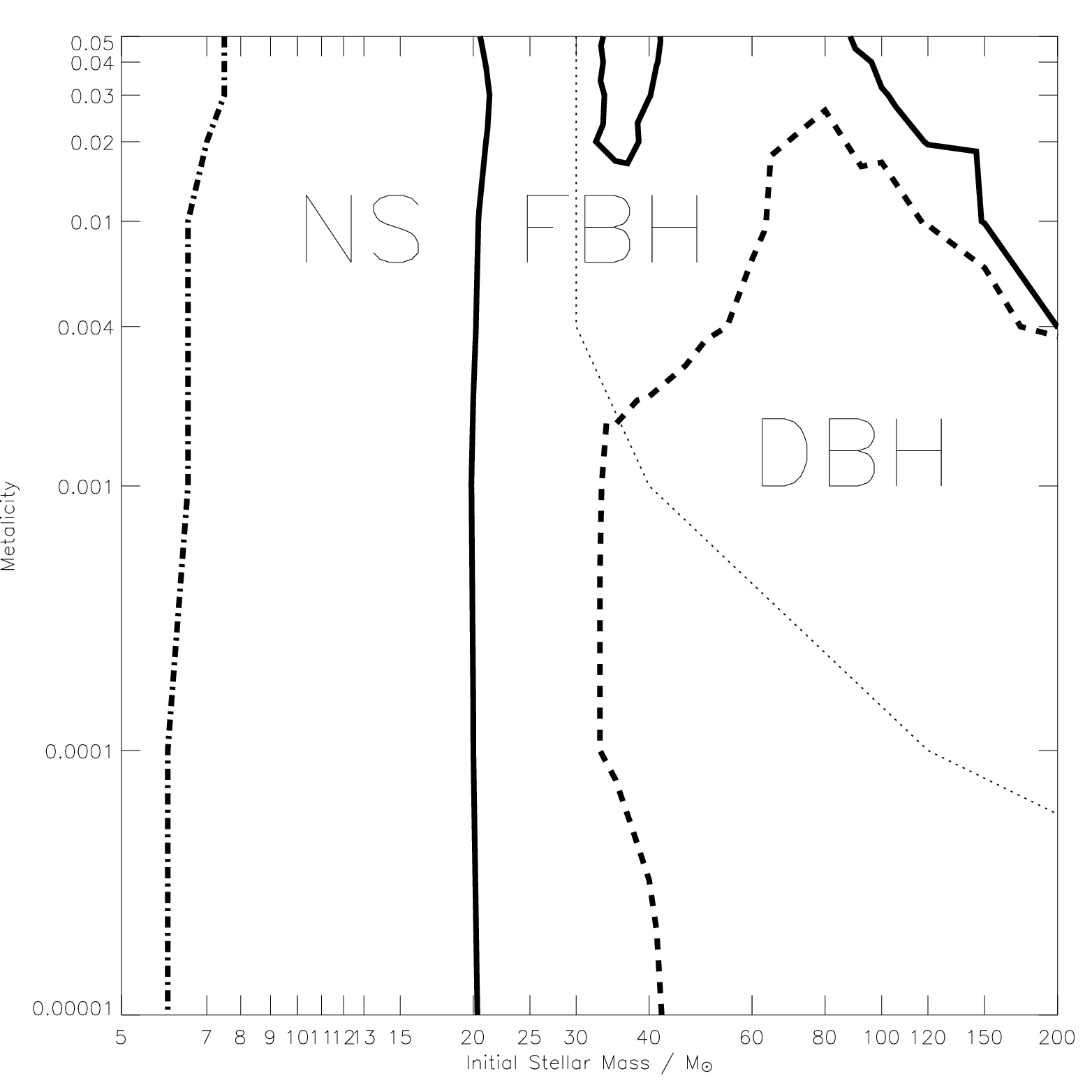}
\caption{He core mass, left no-overshooting, right overshooting. Dot-dashed line - minimum mass for SNe and therefore the boundary between white dwarfs and neutron stars ($M_{\rm core}({\rm He })=1.4M_{\odot}$), solid line boundary between neutron stars and black holes formed by fall back ($M_{\rm core}({\rm He })=8M_{\odot}$), dashed line - boundary between black holes formed by fall back and directly ($M_{\rm core}({\rm He })=15M_{\odot}$). Note this same diagram can be used to determine the visibility of Type Ibc SNe. There are no WR progenitors of $M \le 5M_{\odot}$, but the solid line separates possibly faint SNe and faint SNe, while the dashed line separates faint SNe from SNe with no display. The SNe forming black holes directly are those that have no display. The faint dotted line plots the boundary of type II/I SNe so the number of SNe with no display can be estimated.}
\label{mapD}
\end{figure}

\subsubsection{SNe Type, Figure \ref{mapC}}
Again differences occur with the inclusion of extra mixing. The boundaries for the types of supernovae shift down by around $5-30M_{\odot}$ depending on the metallicity and boundary type. The type IIP/L boundary follows the type II/I boundary although the type IIL region becomes broader at low metallicity. The shape of the type II/I boundary is well defined and is similar between the two diagrams with no type Ibc and IIL supernovae occuring when $Z=10^{-5}$ in our range of initial masses.

With Figures \ref{mapC} and \ref{mapD} and the details of section 3.2 we can identify the mass regime where SN Ibc become unseen. Those SNe that form black holes directly will have no display. Unless a jet driven SNe forms a black hole and produces an observable display. Unseen SNe will affect the ratio of type II/Ibc SNe observed. At low metallicities all type Ibc SNe are unseen. We do not find bright type Ibc SNe in our sample as none of the WR stars have helium cores less than $5M_{\odot}$.

\begin{figure}
\includegraphics[height=80mm]{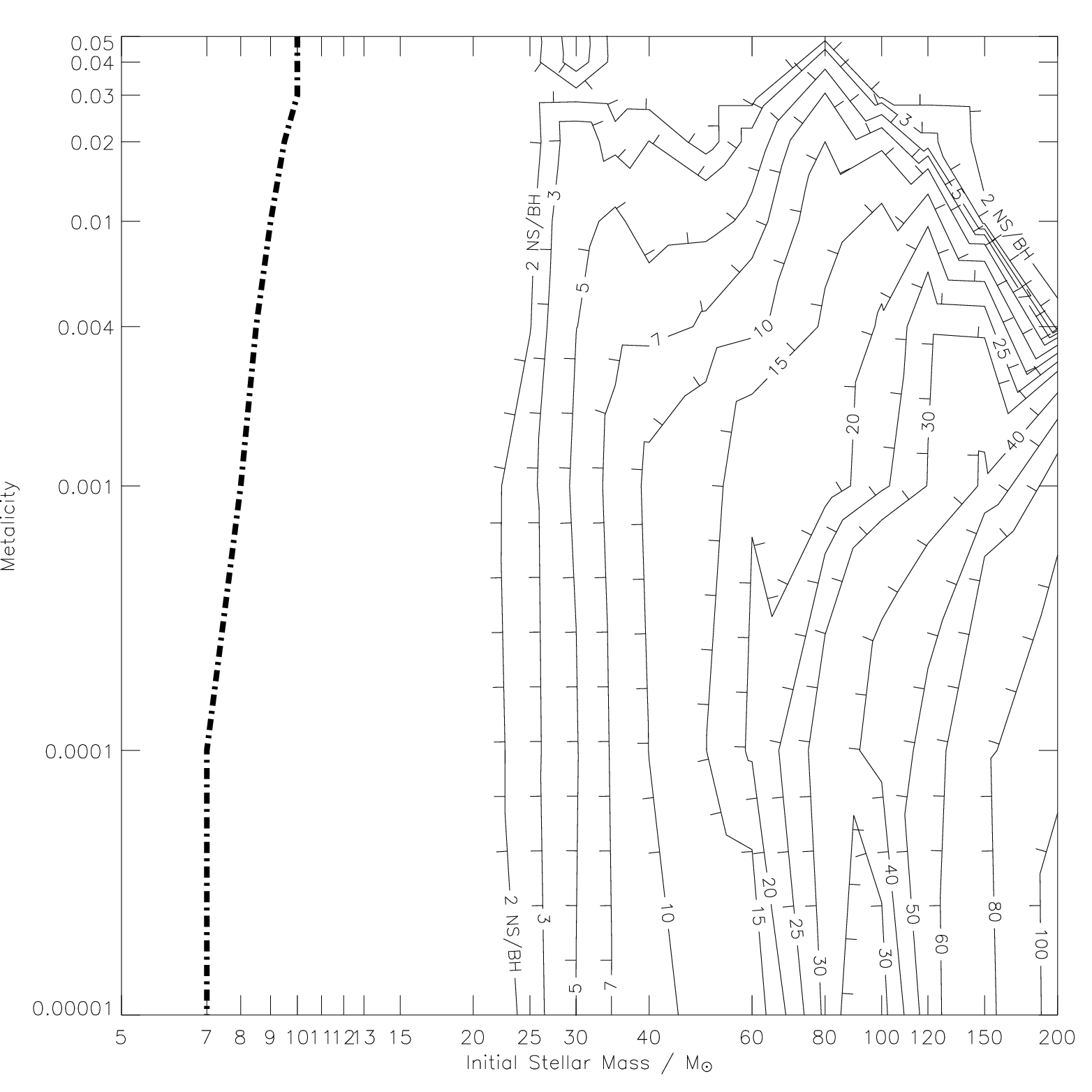}
\includegraphics[height=80mm]{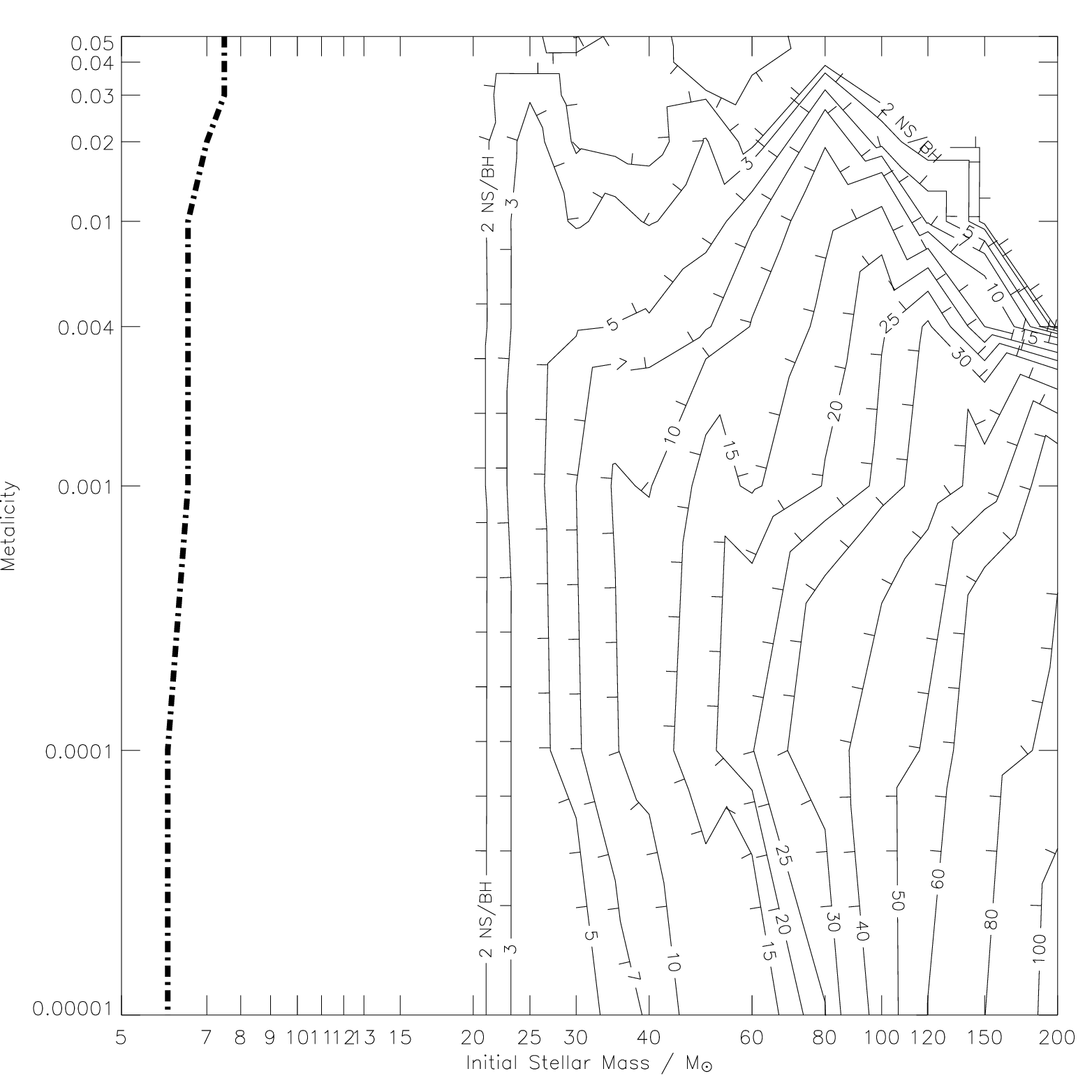}
\caption{Remnant mass by Energy method, left no-overshooting, right overshooting. Units on contours are mass in $M_{\odot}$. Dot-dashed line represents the minimum mass for SNe.}
\label{mapE}
\end{figure}

\subsubsection{SNe Remnant, figures \ref{mapD} \& \ref{mapE}}
In shape our maps are similar to those of \citet{H03} although we do not go to zero metallicity. We also have more structure in the low-mass end of the diagram with the white dwarf to neutron star transition moving to a greater mass at higher metallicity. The change between the no-overshooting and overshooting case is minor with only the transitions moving down in mass while the shape remains the same. Noticeably at high metallicity and mass with overshooting it is possible to see that neutron stars are formed rather than black holes. This is similar to the findings of \citet{H03}. The transitions from neutron star to black hole are at similar positions in \citet{H03} although our value with overshooting is lower by $5M_{\odot}$.

Comparing to the energy method of determining the remnant we find the same transition point between neutron stars to black holes just below $25M_{\odot}$ without overshooting and $21M_{\odot}$ with overshooting. The structure is also quite similar although it is difficult to decide when direct black hole formation may occur rather than fall back on to a black hole. This would require a denser remnant and our $10M_{\odot}$ remnant line seems to follow the He core mass remnant region for direct black holes fairly well. While looking at these diagrams it is sensible to ask where different types of gamma-ray bursts (GRBs) may occur as discussed by \citet{H03}. It is thought that GRBs require a progenitor with a small radius (a few $R_{\odot}$) so the jets that form from the material accreted on to the central black hole can punch their way through the stellar envelope. If the stellar envelope is too large the jets dissipate and lead to a normal SN. Therefore any of our models that loose their hydrogen envelopes to become WR stars and also form a black hole may also be the progenitors of GRBs. It will be interesting to determine how many stars that fit this requirement give rise to an unseen SN compared to those that give a bright display. This could lead to an observational check if it were known how many GRBs have an associated SN.

\section{Discussion}

\subsection{Low-mass SN Progenitors $(M < 25 M_{\odot})$}
Stars with zero-age mass less than $25M_{\odot}$ give rise to red giant SN progenitors and should be the most common single star progenitors observed. The luminosity of these stars determines how many we can expect to find in current observational searches for progenitors. We find the luminosity of these objects depends upon the occurrence of second dredge-up in the late stages of evolution \citep{Smartt2002}. Second dredge-up can increase the final luminosity from $\log (L/L_{\odot}) < 4.6$ to $\log (L/L_{\odot}) > 5.2$. This increases the maximum distance at which the progenitors can be observed. However the details of second dredge-up are complex and depend on the amount of nuclear burning throughout the evolution of the star. The character of stars during second dredge-up depends on initial mass and metallicity and affects the final fate of the star and the material dredged to the surface. We are able to define five classes of red giant progenitors, the masses are for solar metallicity without convective overshooting.

\begin{itemize}
\item $M \le 7M_{\odot}$, stars undergo second dredge-up and thermal pulses with a central CO core as a thermally pulsing AGB star. These lose their envelope and leave CO white dwarfs before before their cores reach $M_{\rm Ch}$.
\item $M \approx 8M_{\odot}$, during or after second dredge-up carbon ignition occurs in a shell because the core is degenerate and has a temperature inversion caused by neutrino losses. The star then undergoes thermal pulses with an ONe core. Since the core mass after dredge-up is less than $M_{\rm Ch}$ we assume these stars lose their envelope forming ONeWDs. These are Super-AGB stars.
\item $M \approx 9M_{\odot}$, carbon is ignited before second dredge-up (in a shell if the centre is degenerate). Thus at dredge-up there is a growing ONe core. If this can reach $M_{\rm Ch}$ before the envelope is lost then the star undergoes a SN. The outcome will depend on the nature of the thermal pulses.
\item $M \approx 10M_{\odot}$, the CO core is greater than $M_{\rm Ch}$ before dredge-up. However shell carbon burning (enhanced by a thin neon burning shell in the most massive stars of this type) drives a convection zone that reduces the size of the CO core to $M_{\rm Ch}$ so dredge-up can occur. This CO material is mixed with the envelope and increases the CO abundance at the surface during second dredge-up. After dredge-up the star has an ONe core of $M_{\rm Ch}$ which progresses to a SN by electron capture by ${\rm Mg}^{24}$.
\item $M \ge 11M_{\odot}$, the helium core or CO core masses are too great for dredge-up to occur. The limiting mass for the helium core is $3M_{\odot}$ and for the CO core $1.5M_{\odot}$. Nuclear burning in these stars progresses until it cannot support the core and a SN occurs.
\end{itemize}

The actual masses vary with metallicity and convective overshooting. As metallicity increases or decreases the transitions increase or decrease respectively by $2M_{\odot}$ at most. Convective overshooting has a strong effect of lowering the masses by $3M_{\odot}$. At solar metallicity the maximum mass for a star to undergo second dredge-up is $8M_{\odot}$. With convective overshooting the above five cases occur in the following mass ranges, $M \le 5M_{\odot}$, $6M_{\odot} \le M \le 7M_{\odot}$, $M \approx 7.5M_{\odot}$, $M \approx 8M_{\odot}$ and $M \ge 9M_{\odot}$

Another intriguing result of convective overshooting is that carbon burning ignites in smaller CO cores so, after second dredge-up, more of the stars have an ONe core smaller than $M_{\rm Ch}$ when thermal pulses begin and therefore produce more ONeWDs than without convective overshooting.

There are few sets of models to which we can compare ours. \citet{IBENX}, and references therein, detail similar models at solar metallicity for this range of masses. Compared to their models our mass limits are lower by a solar mass. \citet{podsi03} also detail stars that undergo second dredge-up and the implications for SNe and find similar limits higher that ours by a solar mass. This change is due to our updated opacity tables and nuclear reaction rates as our models with older tables produce limits that are higher by a solar mass.

\subsection{High-mass SN Progenitors $(M > 25 M_{\odot})$}
The change between the types of SNe II/Ibc is abrupt. However it is more difficult to differentiate between the sub types of Ibc or IIP/L. This is mainly due to the arbitrary nature of determining the way composition affects the SN dynamics. Our results have boundaries that vary less than previous authors because we apply the JNH mass-loss rates. We believe this is appropriate for single stars that are not rapidly rotating or in binaries. 

The problems encountered scaling mass-loss at low metallicity, $Z \le 10^{-4}$, mean that our SNe type boundaries become uncertain. Although the theoretical rates of \citet{KD2002} indicate that a change in the mass-loss rates in this region is expected. This agrees with the cut-off found in our maps.

\subsection{Convolution with an IMF - SNe Ratios.}
Observations tell us that in S0a-Sb galaxies, the ratio of Type II supernovae over Type Ibc supernovae is between 1.2 and 12. In Sbc-Sd galaxies, it is between 2.1 and 16 \citet{Capp1997}. At solar metallicity we find the ratio without overshooting to be 6.5, and with overshooting to be 6.1 for a simple Saltpeter IMF. This is slightly higher than the value of 5 calculated by \citet{H03}. Generally the theory agrees with observation but this basic picture must be altered at lower metallicities because most type Ibc SNe become faint or have no display thereby raising this ratio. However binary systems are likely to dominate Ibc supernova production at low metallicity and this complicates prediction of the SNe type ratio.

\subsection{Remnant Initial Mass Population.}
\begin{figure}
\includegraphics[height=80mm]{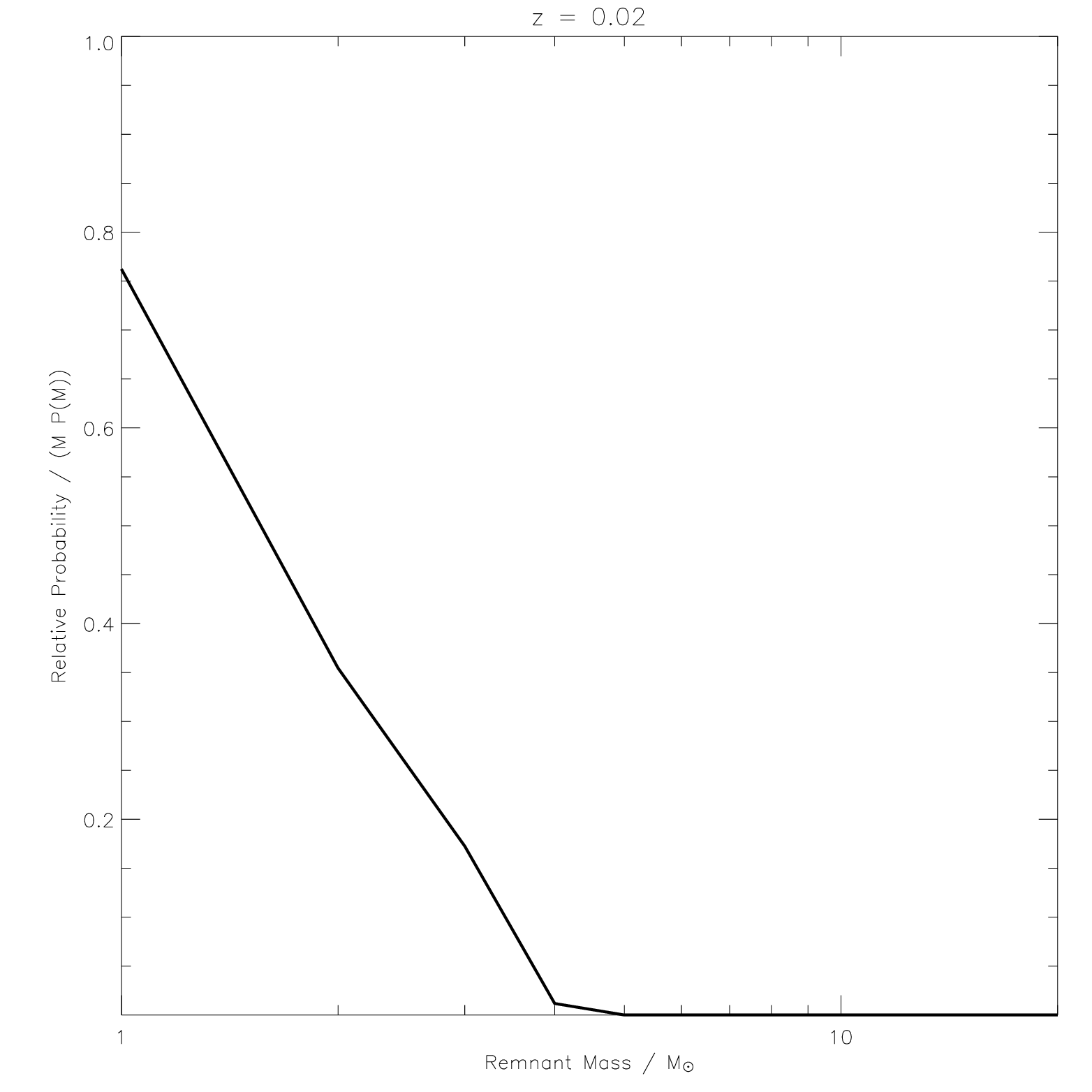}
\includegraphics[height=80mm]{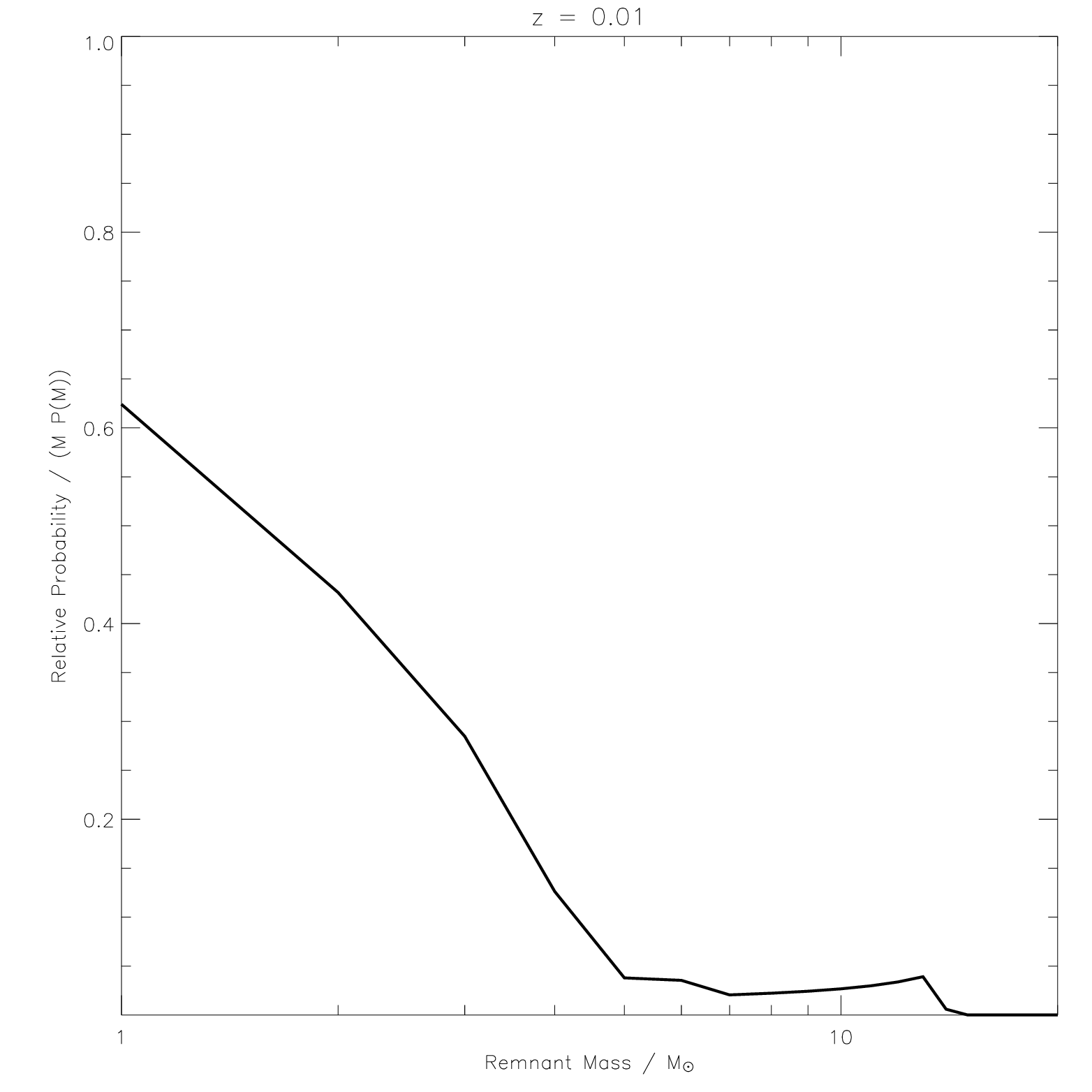}
\includegraphics[height=80mm]{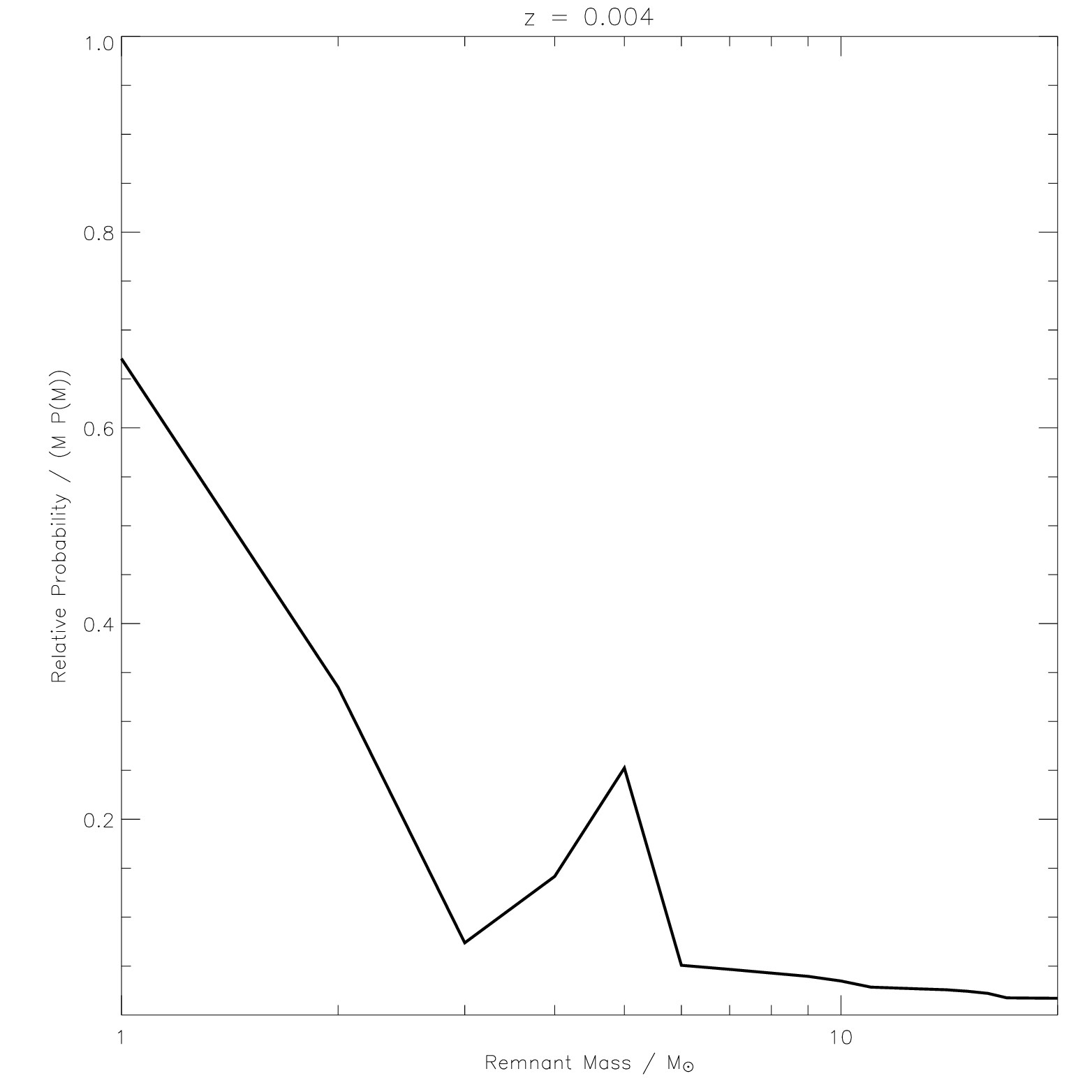}
\includegraphics[height=80mm]{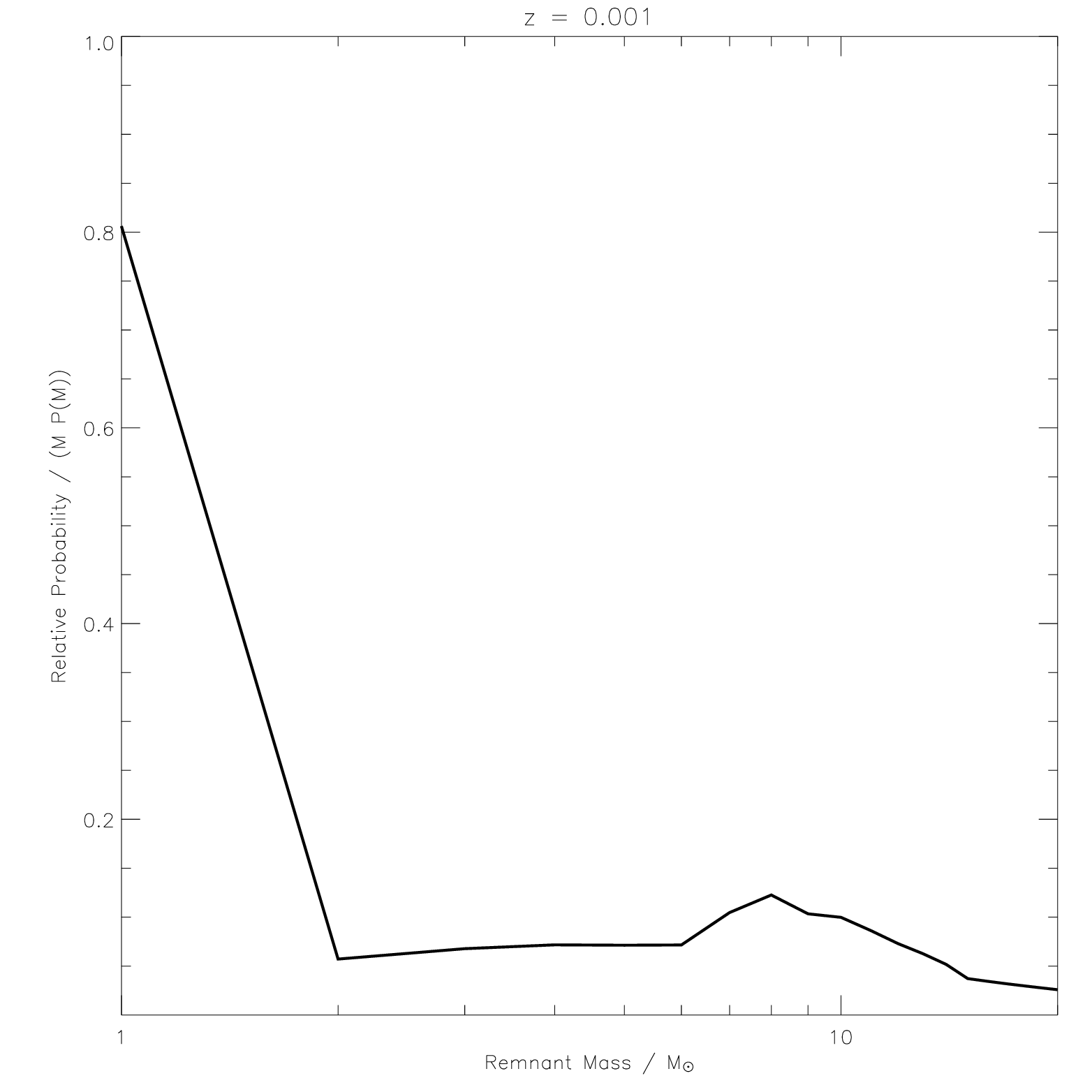}
\caption{Graphs of probability of remnant masses obtained by convolution of the remnant mass diagram with overshooting and a Salpeter IMF. The value at the top of each individual graph is the metallicity plotted. The probability has been divided by the mass of remnant. There are sharp cut-offs at lower metallicity and the distributions become smoother at lower metallicity.}
\label{gridimf}
\end{figure}

Figure \ref{gridimf} shows the probability of a remnant having a mass $M_{\rm rem}$. At the highest metallicities the remnants are all of very low-mass due to a maximum remnant mass cut off, with neutron stars dominant. Then as metallicity decreases the maximum remnant mass increases until the distribution becomes smooth with no maximum mass cut off. There is also a peak above the typical minimum mass for a black hole. These are the most common black holes at that metallicity. Although as the remnant mass used for these diagrams comes from a very simple calculation they are at best estimates of the distribution.

Very massive $(M \ge 100M_{\odot})$ black holes can form in SN however this can only happen at low metallicity ($Z < 10^{-3}$). The population of such black holes will be very small as they are only formed in the most massive stars. At higher metallicities massive black holes are not formed as mass loss limits the size of the SN progenitor and thus the size of the remnant. Any black holes observed to be very massive in high metallicity environments must therefore have grown via accretion from a binary companion or must have formed from an an earlier population of stars.

\subsection{The Progenitor of 2003gd and other SNe.}
\begin{figure}
\includegraphics[height=80mm,angle=270]{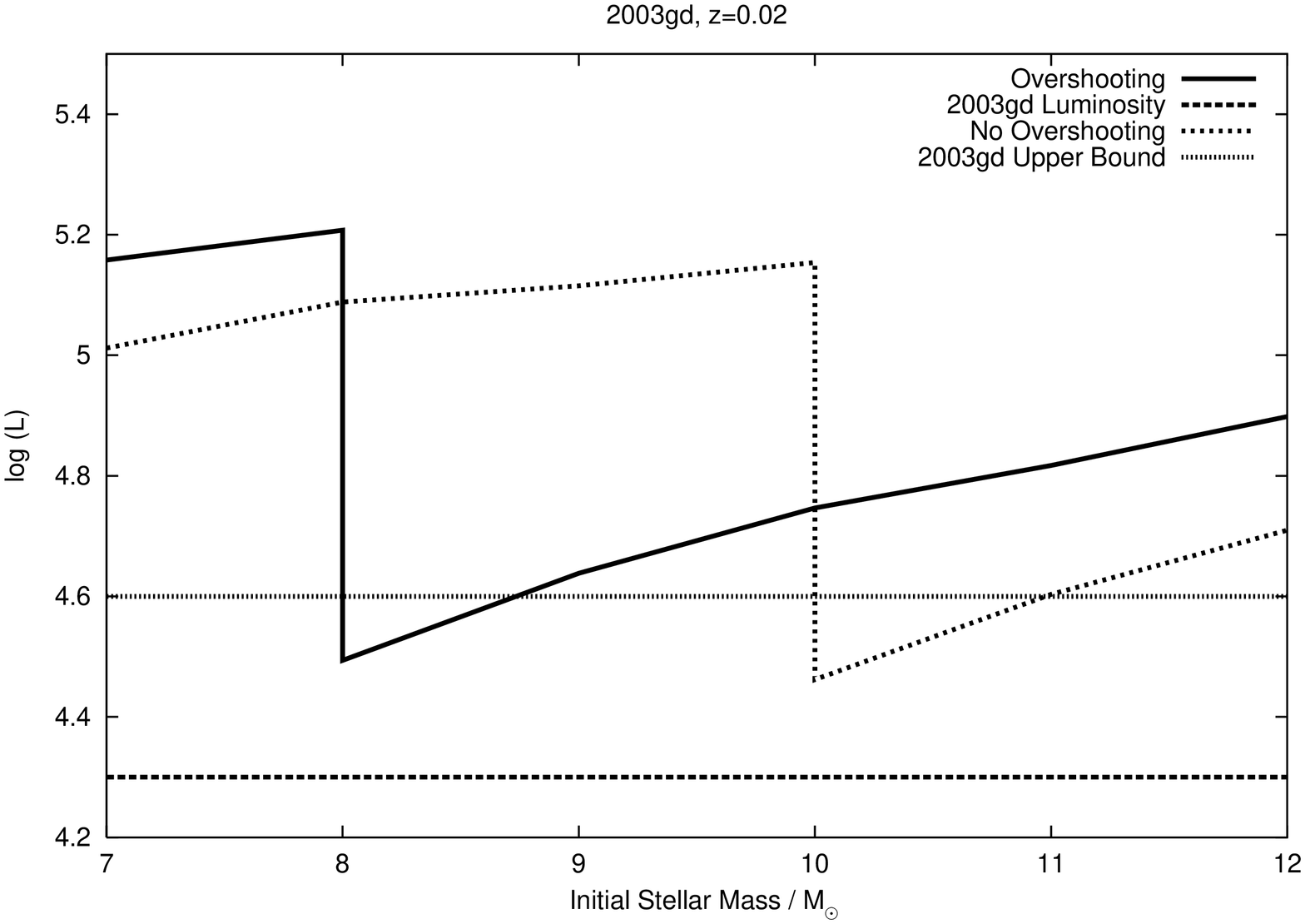}
\includegraphics[height=80mm,angle=270]{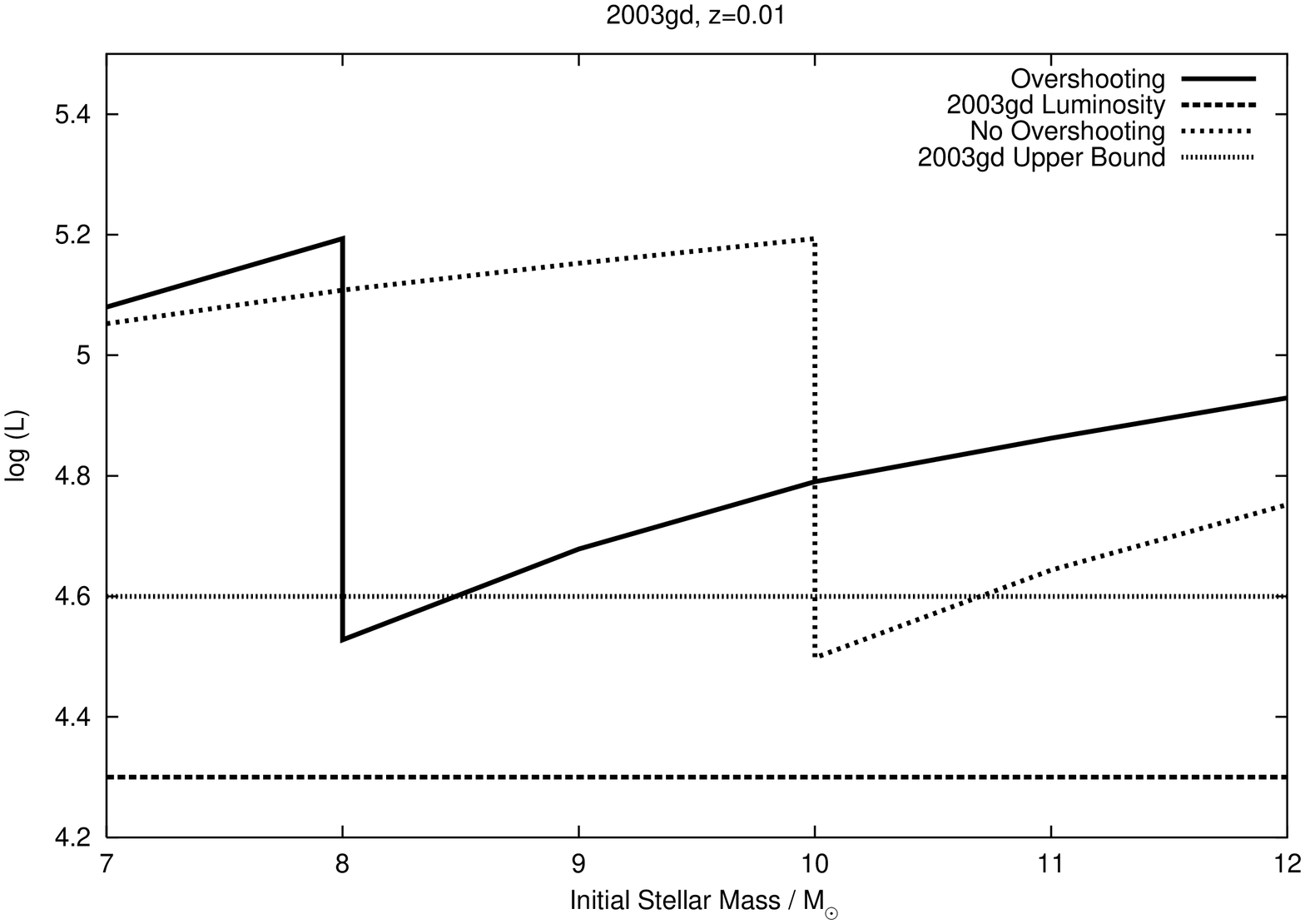}
\caption{Comparing observations of 2003gd with models at solar and half solar metallicity. Both cased with or without overshooting fit the observations however the mass range is tightly constrained in both cases.}
\label{2003gd}
\end{figure}

The progenitor of SN2003gd is the first red progenitor to be observed. The observed luminosity is such that $\log_{10}(L/L_{\odot}) = 4.3 \pm 0.3$ \citep{SJM03}. This places the progenitor in the very faintest region of our maps. Figure \ref{2003gd} shows the luminosity limits of 2003gd compared with the model luminosities at solar and half solar metallicity. The star cannot have gone through second dredge-up. This places important constraints on the mass of the progenitor. However the luminosities of the candidate stars are both near to the upper limit of the observation so we cannot distinguish between the overshooting or non-overshooting case. The minimum mass we can have for this star is $8M_{\odot}$. 

\begin{figure}
\includegraphics[height=80mm]{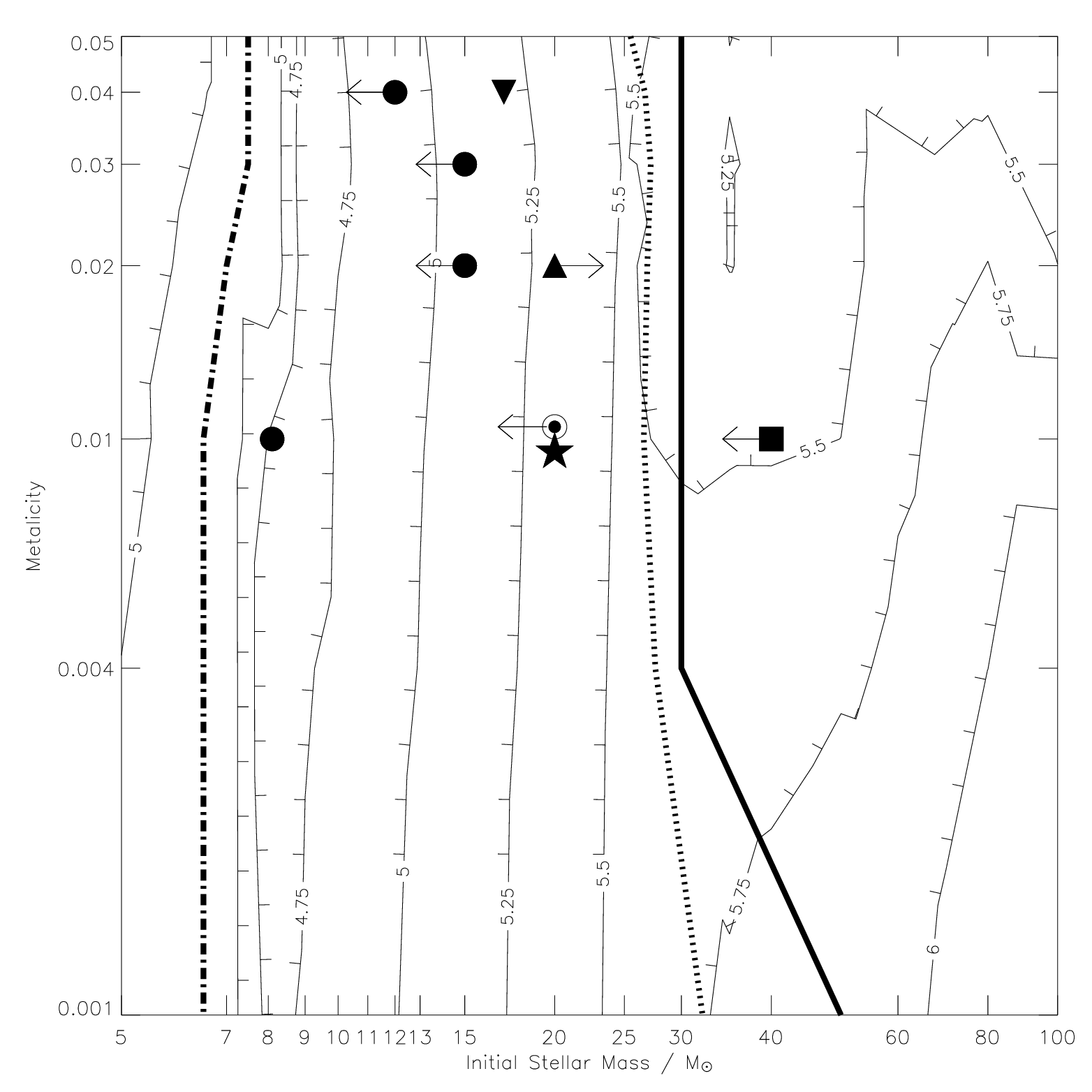}
\caption{Comparing values from maps to observations of progenitors. The contours are as in Figure 2 while the thick lines are from Figure 4. The map is with overshooting and zoomed in on the region where observations exist. The data points are taken from \citet{S03a}. Filled circles represent type IIP SNe, the circle with filled centre represents the type IIL SN, the triangle represents the type IIn SN, the inverted triangle represents the type IIb SN, the star symbol represents the II-pec SN and the square represents the type Ic SN.}
\label{zoomed}
\end{figure}

\citet{S03a} compiled details on the sample of well studied SN progenitors (shown in Table \ref{ptable} and Figure \ref{zoomed}). We can calculate various details to check against these observations. Most match well with our overshooting models although the various types of type II supernovae do not match our IIP predictions. This may be due these stars being binaries for example recent observations \citep{Maund04} have shown the progenitor to 1993J was part of a binary. There are also strong arguments that 1987A was part of a binary too \citep{podsi92}. Except for 1993J the ejected masses also match up well to those calculated from observations. As this list grows in length and detail more tests against our maps will become possible.

\begin{table}
\caption{Comparing values from maps to observations of progenitors. Adapted from table in \citet{S03a}.}
\label{ptable}
\begin{tabular}{lcccccccccc}
\hline
     & Observed &         &Observed&     & Observed &Map  &Map  &Map        &Map&Map\\
     &   SNe & Observed&Stellar& Initial  & Ejected     &SNe  &He Core&Energy   &Ejected&Stellar\\
SN   &  Type   & $Z/Z_{\odot}$&Type &Mass $/M_{\odot}$& Mass $/M_{\odot}$&Type&Remnant&Remnant&Mass/$M_{\odot}$&Type\\
\hline
2002ap & Ic     & 0.5     & WR?     &	$<40$ &2.5-5 &Ibc  &BH   &$3M_{\odot}$ BH&  7&WR\\
1997bs & IIn    & ~1 & ?       &	$>20$ & -     &IIP &NS  &NS        & 10 & M\\
1987A  & II-pec & 0.5     & B3Ia    &  $20$ &15    &IIP  &NS   &NS	   & 15 & K  \\
1993J  & IIb    & 2       & K &  $17$ &3     &IIP  &NS   &NS	   & 12.5 & M\\
1980K  & IIL    & 0.5     & ?       &	$<20$ &  -    &IIP  &NS   &NS	   & 16 & K \\
2003gd & IIP    & 0.5     & M & $8$ & - & IIP &NS&NS& 6 &M\\
2001du & IIP    & ~1 & G-M     &	$<15$ &-     &IIP  &NS   &NS	   & 12 & M \\
1999em & IIP    & 1-2     & K-M     &	$<15$ &5-18  &IIP  &NS   &NS	   & 12 & M \\
1999gi & IIP    & ~2 & G-M     &	$<12$ &10-30 &IIP  &NS   &NS	   & 9 & M  \\
\hline
\end{tabular}
\end{table} 

\section{Conclusion}
We find that our results generally agree with \citet{H03}, although our maps vary less with metallicity because we apply JNH mass-loss rates rather than NJ over most of the H-R diagram except for a small region at low metallicity and large initial mass. Observational evidence does match well with our maps and we can also point out that the most common type of progenitors, red giants, should be like 2003gd, faint compared to the rarer more massive stars. Therefore observations may find it difficult to find red progenitors unless some of those that undergo 2nd dredge-up do evolve to SN. In which case there will be a population of very luminous red progenitors that have not been observed to date.

\section{Acknowledgements}
We would like to thank the referee Alexander Heger for his helpful comments, Stephen Smartt for an observers view and Philipp Podsiadlowski for helpful comments. JJE would like to thank PPARC for his funding. CAT would like to thank Churchill College for his Fellowship.

\bsp

\newpage

\newpage

\begin{thebibliography}{99}
\bibitem[\protect\citeauthoryear{Angulo et al}{1999}]{NACRE}Angulo C. et al., Nucl. Phys. A656 (1999)3-187
\bibitem[\protect\citeauthoryear{B\"ohm-Vitense}{1958}]{BV58}B\"ohm-Vitense E, 1958, ZA, 46, 108
\bibitem[\protect\citeauthoryear{Cappellaro et al.}{1997}]{Capp1997} Cappellaro E., Turatto M., Tsvetkov D. Yu., Bartunov O.S., Pollas C., Evans R., Hamuy M., 1997, A\&A, 322, 431
\bibitem[\protect\citeauthoryear{Caughlan \& Fowler}{1988}]{CF1988}Caughlan G.R., Fowler W.A., 1988, ADNDT, 40, 283
\bibitem[\protect\citeauthoryear{Crowther}{2001}]{Crow2001}Crowther P.A., in The influence of binaries on stellar population studies, Dordrecht: Kluwer Academic Publishers, 2001
\bibitem[\protect\citeauthoryear{Dray \& Tout}{2003}]{DT03} Dray L.M., Tout C.A., 2003, MNRAS, 341, 299
\bibitem[\protect\citeauthoryear{Dray}{2003}]{DrayThesis}Dray L., 2003, Phd Disertation, University of Cambridge.
\bibitem[\protect\citeauthoryear{Eggleton}{1971}]{E71} Eggleton P.P., 1971, MNRAS, 151, 351
\bibitem[\protect\citeauthoryear{Eldridge \& Tout}{2004}]{E03} Eldridge J.J., Tout C.A., 2004, MNRAS, 348, 201
\bibitem[\protect\citeauthoryear{Ensman \& Woosley}{1988}]{EW88} Ensman L.M., Woosley S.E., 1988, ApJ, 333, 754
\bibitem[\protect\citeauthoryear{Fryer}{1999}]{Fry99}Fryer C.L., 1999, ApJ, 522, 413
\bibitem[\protect\citeauthoryear{Hamann \& Koesterke}{1998}]{HKO98}Hamann W.-R., Koesterke L., 1998, A\&A, 335, 1003
\bibitem[\protect\citeauthoryear{Heger \& Langer}{1996}]{HL96}Heger, A., Langer, N., 1996, A\&A, 315, 421
\bibitem[\protect\citeauthoryear{Heger et al.}{2003}]{H03}Heger A., Fryer C.L., Woosley S.E., Langer N., Hartmann D.H., 2003, 
\bibitem[\protect\citeauthoryear{Hurley, Pols \& Tout}{2000}]{HPT00}Hurley J.R., Pols O.R., Tout C.A., 2000, MNRAS, 315,543
\bibitem[\protect\citeauthoryear{Iglesias \& Rogers}{1996}]{IR96} Iglesias C.A., Rogers F.J., 1996, ApJ, 464, 943
\bibitem[\protect\citeauthoryear{de Jager, Nieuwenhuijzen \& van der Hucht}{1988}]{dJ}de Jager C., Nieuwenhuijzen H., van der Hucht K.A., 1998, A\&ASS, 72, 259
\bibitem[\protect\citeauthoryear{Kudritzki}{2002}]{KD2002} Kudritzki R.P., 2002, ApJ, 577, 389 
\bibitem[\protect\citeauthoryear{Langer}{1989}]{Langer}Langer, N., 1989, A\&A, 220, 135
\bibitem[\protect\citeauthoryear{MacFadyen, Woosley \& Heger}{2001}]{MWH01}MacFadyen A.I., Woosley S.E., Heger A., 2001, ApJ, 550, 410
\bibitem[\protect\citeauthoryear{Maund et al.}{2004}]{Maund04}Maund J.R., Smartt S.J., Kudritzki R.P., Podsiadlowski Ph., Gilmore G.F., 2004, Nature, 427, 129
\bibitem[\protect\citeauthoryear{Meynet et al}{1994}]{MM}Meynet G., Maeder A., Schaller G., Schaerer D, Charbonnel C., 1994, A\&AS, 103, 97
\bibitem[\protect\citeauthoryear{Nieuwenhuijzen \& de Jager}{1990}]{NJ90}Nieuwenhuijzen H., de Jager C., 1990, A\&A, 231, 134
\bibitem[\protect\citeauthoryear{Nugis \& Lamers}{2000}]{NL00}Nugis T., Lamers H.J.G.L.M., 2000, A\&AS, 360, 227
\bibitem[\protect\citeauthoryear{Pinsonneault}{1997}]{b7b}Pinsonneault M., 1997, ARA\&A, 35, 557
\bibitem[\protect\citeauthoryear{Podsiadlowski}{1992}]{podsi92} Podsiadlowski P., 1992, PASP, 104, 717
\bibitem[\protect\citeauthoryear{Podsiadlowski et al.}{2003}]{podsi03} Podsiadlowski P., Langer N., Poelarends A.J.T, Rappaport S., Heger A., Pfahl E., 2003, astro-ph/0309588
\bibitem[\protect\citeauthoryear{Pols et al.}{1995}]{P95} Pols O.R., Tout C.A., Eggleton P.P., Han Z., 1995, MNRAS, 274, 964
\bibitem[\protect\citeauthoryear{Pols et al.}{1997}]{P97} Pols O.R., Tout C.A., Schr\"oder K.-P., Eggleton P.P., Manners J., 1997, MNRAS,289,869
\bibitem[\protect\citeauthoryear{Pols et al.}{1998}]{P98} Pols O.R., Schr\"oder K.-P., Hurley J.R., Tout C.A., Eggleton P.P., 1998, MNRAS, 298,525
\bibitem[\protect\citeauthoryear{Ritossa, Garcia-Berro \& Iben}{1999}]{IBENX}Ritossa C., Garcia-Berro E., Iben I., 1999, ApJ, 515, 381
\bibitem[\protect\citeauthoryear{Schr\"oder, Pols \& Eggleton}{1997}]{SPE97}Schr\"oder, K.P., Pols, O.R., Eggleton, P.P., 1997, MNRAS, 285, 696
\bibitem[\protect\citeauthoryear{Smartt et al.}{2002}]{Smartt2002}Smartt S.J., Gilmore G.F., Tout C.A., Hodgkin S.T.,  2002, ApJ, 565, 1089
\bibitem[\protect\citeauthoryear{Smartt et al}{2003}]{S03a}Smartt S.J., Maund J.R., Gilmore G.F., Tout C.A., Kilkenny D., Benetti S., 2003, 2003, MNRAS, 343, 735
\bibitem[\protect\citeauthoryear{Smartt el al.}{2004}]{SJM03}Smartt S.J., Maund J.R., Hendry M.A., Tout C.A., Gilmore G.F., Mattila S., Benn C.R., 2004, Science, 303, 499
\bibitem[\protect\citeauthoryear{Stephenson \& Green}{2002}]{SG02}Stephenson F.R., Green D.A., Historical supernovae and their remnants. International series in astronomy and astrophysics, vol. 5. Oxford: Clarendon Press, 2002, ISBN 0198507666
\bibitem[\protect\citeauthoryear{Van Dyk et al.}{2003}]{VD03} Van Dyk S.D., Li, W., Filippenko A.V., 2003, PASP, 115, 1289
\bibitem[\protect\citeauthoryear{Vink, de Koter \& Lamers}{2000}]{VKL2000}Vink J.S., de Koter A., Lamers H.J.G.L.M., 2000, A\&A, 362, 295
\bibitem[\protect\citeauthoryear{Vink, de Koter \& Lamers}{2001}]{VKL2001}Vink J.S., de Koter A., Lamers H.J.G.L.M., 2001, A\&A, 369, 574
\bibitem[\protect\citeauthoryear{Wellstein \& Langer}{1999}]{WL1999} Wellstein S., Langer N., 1999, A\&A, 350, 148
\bibitem[\protect\citeauthoryear{Woosley, Heger \& Weaver}{2002}]{WHW02}Woosley S.E., Heger A., Weaver T.A., 2002, Rev..
\label{lastpage}
\end{thebibliography}
\end{document}